# Geomorphologic Mapping of Titan's Polar Terrains: Constraining Surface Processes and Landscape Evolution

S.P.D. Birch*[1], A.G. Hayes[1,2], W.E. Dietrich[3], A.D. Howard[4], C.S. Bristow[5], M.J. Malaska[6], J.M. Moore[7], M. Mastrogiuseppe[2], J.D. Hofgartner[2,6], D.A. Williams[8], O.L. White[7], J.M. Soderblom[9], J.W. Barnes[10], E.P. Turtle[11], J.I. Lunine[2], C.A. Wood[12], C.D. Neish[13], R.L. Kirk[14], E.R. Stofan[15], R.D. Lorenz[10], and R.M.C. Lopes[6]

[1] Department of Earth and Atmospheric Sciences, Cornell University, Ithaca NY, USA

[2] Department of Astronomy, Cornell University, Ithaca NY, USA

[3] Department of Earth and Planetary Science, University of California Berkeley, Berkeley CA, USA

[4] Department of Environmental Sciences, University of Virginia, Charlottesville VA, USA

[5] Department of Earth and Planetary Sciences, Birkbeck University of London, London, U.K.

[6] Jet Propulsion Laboratory / California Institute of Technology, Pasadena CA, USA

[7] Space Sciences Division, NASA Ames Research Facility, Moffett Field CA, USA

[8] School of Earth and Space Exploration, Arizona State University, Tempe, AZ, USA

[9] Department of Earth, Atmospheric and Planetary Sciences, Massachusetts Institute of Technology, Boston MA, USA

[10] Department of Physics, University of Idaho, Moscow ID, USA

[11] John Hopkins University Applied Physics Lab, Laurel MD, USA

[12] Planetary Science Institute Tucson, Tucson AZ, USA

[13] University of Western Ontario, London ON, Canada

[14] US Geological Survey, Astrogeology Division, Flagstaff AZ, USA

[15] Department of Earth and Planetary Science, University College London, UK

# ABSTRACT

We present a geomorphologic map of Titan's polar terrains. The map was generated from a combination of Cassini Synthetic Aperture Radar (SAR) and Imaging Science Subsystem imaging products, as well as altimetry, SARTopo and radargrammetry topographic datasets. In combining imagery with topographic data, our geomorphologic map reveals a stratigraphic sequence from which we infer process interactions between units. In mapping both polar regions with the same geomorphologic units, we conclude that processes that formed the terrains of the north polar region also acted to form the landscape we observe at the south. Uniform, SAR-dark plains are interpreted as sedimentary deposits, and are bounded by moderately dissected uplands. These plains contain the highest density of filled and empty lake depressions, and canyons. These units unconformably overlay a basement rock that outcrops as mountains and SAR-bright dissected terrains at various elevations across both poles. All these units are then superposed by surficial units that slope towards the seas, suggestive of subsequent overland transport of sediment. From estimates of the depths of the embedded empty depressions and canyons that drain into the seas, the SAR-dark plains must be >600 m thick in places, though the thickness may vary across the poles. At the lowest elevations of each polar region, there are large seas, which are currently liquid methane/ethane filled at the north and empty at the south. The large plains deposits and the surrounding hillslopes may represent remnant landforms that are a result of previously vast polar oceans, where larger liquid bodies may have allowed for a sustained accumulation of soluble and insoluble sediments, potentially forming layered sedimentary deposits. Coupled with vertical crustal movements, the resulting layers would be of varying solubilities and erosional resistances, allowing formation of the complex landscape that we observe today.

# 1. INTRODUCTION

Titan's lakes, seas and surrounding hillslopes contain vast amounts of information regarding the history and evolution of Saturn's largest moon. With an atmospheric pressure at the surface of 1.5 bars and a surface temperature of 91-95 K, methane and ethane are both able to condense out of the atmosphere and rain to the surface (Atreya 2006), where the fluid runoff concentrates, incises channels and transports sediment (Collins, 2005; Burr et al., 2006). Landforms common to Earth are found across Titan, and include lakes and seas (Stofan et al., 2007; Hayes et al., 2008), river valleys (Lorenz et al., 2008; Burr et al., 2013), fans and deltas (Witek et al., 2015; Radebaugh et al., 2016; Birch et al., 2016) and mountains (Barnes et al., 2007; Radebaugh et al., 2007; Cook-Hallett et al., 2015; Liu et al., 2016). Yet under Titan conditions, these familiar landforms have all formed and evolved under vastly different environmental and physical conditions from Earth.

Of particular interest to this study are the fluvial and lacustrine features clustered in Titan's polar regions. Flybys of Titan by the Cassini spacecraft have revealed numerous small lakes and vast seas at the north (Stofan et al., 2007; Hayes et al., 2008), while the south shows a markedly different environment with only a few lakes (Aharonson et al., 2013). The mechanisms and temporal variations in the development of Titan's landscapes can be revealed through a process-based study of the relationships among their morphologies. For example, the shapes of hillslope profiles (e.g., convex vs. concave) suggest processes that both produce and drive mass transport of sediment across the landscape (e.g., Gilbert 1877; Dietrich et al., 2003). The occurrence of channelized flow requires mechanisms that induce concentrated erosion and transport of sediment, resulting in both erosional valley networks (Burr et al., 2013) and depositional deltas and fan-like features (Wall et al., 2010; Witek et., 2015; Radebaugh et al., 2016; Birch et al., 2016).

The identification of the dominant geomorphologic processes in various locations across Titan's polar regions provides insight into the history of the landscape across both time and space (e.g., Dietrich et al., 2003; Grotzinger et al., 2013). We identify these processes through the construction of a geomorphologic map, which allows us to reduce the surface into a collection of units distinguished by the processes responsible for their formation and evolution (Bloom, 1991).

The first part of this paper presents the first, most complete geomorphologic map of Titan's polar regions through the analysis of Synthetic Aperture Radar (SAR), Imaging Science Subsystem (ISS) and altimetry data. Our work expands upon previous studies by coupling the two-dimensional images with topographic data, using the topographic information as an inherent characteristic of our mapping units. These geomorphologic maps differ from geologic maps, as geologic mapping identifies the spatial boundaries of different rock types and surficial deposits, as well as structural features such as faults, bedding orientations and folds. Geomorphologic mapping focuses on surface morphology and topography, and what they reveal about landscape evolution.

In the second part of our work, we use the relationships between the distribution and appearance of our mapped geomorphologic units to investigate the sequence of events that led to their formation. Our process-based approach allows us to develop an interpretive model for the evolution of Titan's polar terrains. We suggest a model in which large volumes of sediment were deposited within larger polar liquid bodies in a previous geologic epoch. The present-day landscape may thus be an erosional remnant that is being lowered in elevation through time.

# 2. MAPPING METHODOLOGY

Our geomorphologic map is centered about Titan's polar regions, and includes latitudes greater than 60° (Fig. 1). We use a combination of the Cassini SAR images (Elachi et al., 2004) along with topographic data in the form of SARTopo (Stiles et al., 2009), altimetry (Zebker et al., 2009) and sparsely distributed Digital Terrain Models (DTMs: Kirk et al., 2012). All our mapping is carried out using the ArcGIS™ cartography software where we were able to combine individual, rasterized SAR swaths (up to and including data from T120) with the topographic information, using a polar stereographic projection. Mapping uses the incidence angle corrected SAR swaths so as to minimize any geometric variations (Farr, 1993), giving us as consistent a dataset as possible. After defining our geomorphologic units, mapping is conducted systematically at a scale of 1:300,000.

　　Figure 1 details the data we use to generate our maps, while Figs. 2 and 3 show the resulting distribution of geomorphologic units. The SARTopo, altimetry and DTM data in Fig. 1 are essential to our mapping efforts, as their inclusion allows us to map the three-dimensional form of the landscape. Absolute uncertainties in elevation estimates for the topographic data range from 8 meters for the altimeter over liquid bodies and 35 m over solid surfaces (e.g., Poggiali et al., in prep) to hundreds of meters for the DTMs (Kirk et al., 2012), making estimates of absolute elevations across length scales spanning the entire pole inaccurate. The topography along a single SARTopo swath gives a reliable topographic profile, with ~75 m vertical resolution (Stiles et al., 2009) that matches both DTM and altimetry profiles, but ephemeris errors between different swaths make the determination of absolute elevations between swaths less reliable (e.g., Hayes et al., submitted). Instead, focus is placed on the *relative* elevations between morphologically distinct units within local regions of the pole. In locations where topographic data are not available or are insufficient in resolution, SOCETSET photogrammetric software is used (Miller and Walker, 1993). Because the generation of large DTMs (e.g., Kirk et al., 2012) is beyond the scope of this work, we instead use the software to infer relative topographic relationships from viewing the area in stereo.

　　There are large regions of both poles where topographic data are not available (Fig. 1). We define our units by focusing the initial mapping to the regions with the most topographic coverage. Once a consistent set of observable patterns for individual units can be mapped within the selected regions, mapping is completed for the remainder of each pole. This technique is a standard practice in producing geomorphologic maps (e.g., Williams et al., 2002). In regions where there is no SAR coverage, lower resolution HiSAR along with ISS images are used at a lower mapping resolution (dotted units in Figs. 2b and 3b) by correlating these datasets to specific units where high resolution data are available.

　　In areas with a high coverage of topographic data, we subdivide the terrain into fifteen geomorphologic units (see Table 1) according to their morphology, topography, and degree of dissection. Figure 4 details the distinction among units, where we qualitatively plot the units according to both topographic slopes and radar backscatter variability. Regions such as the mountains have high topographic slopes that give rise to their brightness in SAR images. They have a relatively uniform backscatter however, in that features appear similar, independent of viewing geometry. The dissected uplands, on the contrary, are both topographically high and variable in their backscatter.

　　We use a three-letter classification scheme for most of our geomorphologic units, where the first letter defines topography, the second defines radar texture and the final letter designates brightness in SAR images. For example, the SAR-dark, high plains (Section 3.2.1) would appear as $Hu_d$, where they are topographically high ($H$), uniform in backscatter ($u$) and SAR-dark ($d$). Units such as mountains, lake depressions, seas, and alluvial fans use a different scheme. The color scheme is defined in Figs. 2b and 3b used throughout the work.

　　ISS mosaics were overlain on the SAR data to look for further correlations between subunits in regions of particular interest. These relations are also detailed in Table 1 for each unit. ISS can yield important properties of the surface that are transparent at radio wavelengths (e.g., Porco et al., 2004). Because of the different spatial resolutions between ISS and SAR, the use of the near-IR information for all units is not possible. In some cases, we can link morphology to near-IR brightness, while in units like alluvial fans and fluvial valleys, their small spatial scales limit such analyses. Altimetry data in combination with the SAR observations, also provide information regarding possible surface composition (e.g., Wye et al., 2009; Michaelides et al., 2016). The behavior of the radar backscatter, in which we quote the normalized radar backscatter cross-section ($\sigma_0$), at nadir and off-nadir incidence angles reveals whether backscatter variations are due to changes in the surface roughness or if there is a difference in the dielectric constant and thus composition and porosity (Farr et al., 1993). These spectral characteristics are included in Table 1 and aid in further classification and understanding of subunits.

# 3. GEOMORPHOLOGIC UNITS: DESCRIPTIONS
## 3.1 VARIABLY SAR-BRIGHT UNITS
### 3.1.1. Mountains & SAR-Bright, Dissected Terrains

Of the four datasets that we utilize in our work (SAR imagery, topography, altimetry and ISS imagery), the mountains and SAR-bright, dissected terrains differ only in their topographic relief. We therefore subdivide the units in our mapping to express this difference in topography, where the separation in our mapping allows for a better visual illustration of the topographic expression of the landscape. However, we acknowledge that their morphologic formation and emplacement are likely similar, as has been the case for previous mapping studies (e.g., Lopes et al., 2010; Williams et al., 2011; Malaska et al., 2016; Lopes et al., 2016).

3.1.1.a. Mountains (*Mtn*) - **Type Example** – 77º S, 60º W - Fig. 5a/b/c

Mountains are defined as SAR-bright, elevated terrains with an observable bright/dark pairing resulting from either a layover effect or shadowing (e.g., Farr et al., 1993). We define a mountain to have the greatest relief within a chosen drainage basin, where relief is the difference in elevation between the highest and lowest points in a drainage basin. Mountains, on average have a relief of ~400 m above surrounding plains, and can be elevated above surrounding units by up to ~850 m.

We also define the mountains to have a more uniform radar backscatter, where the radar backscatter includes effects from volume scattering, wavelength-sized scatterers and dielectric/material properties (Farr, 1993). Where there are overlapping SAR swaths of the same mountain, the feature has a consistent appearance, with only the bright-dark pairings changing orientations as a result of the high slopes and varying incidence and azimuth angles. The brightness of the unit, we assume, is controlled by large-scale facets or slopes oriented towards the spacecraft, and a largely diffuse scattering behavior at smaller scales. This is illustrated in Fig. 4 where the relationships between slope and SAR-backscatter variance are drawn. The relationships for the remainder of the units are also shown in this figure.

While mountains are rare at the north, they are often observed at the south, particularly around the region of Ontario Lacus (Wall et al., 2010). These mountains are shown in Fig. 6a, and appear to enclose Ontario Lacus within a basin of its own. The majority of these mountains align nearly parallel to Ontario Lacus. The remainder of the mountains in the region appears to align along preferred directions, nearly parallel to the long channels draining into the Ontario Lacus basin. Indeed, throughout the south, mountains appear elongated along given directions (Cook-Hallett et al., 2015; Fig. 2b).

In ISS images, the mountains also have a low near-IR brightness at the poles. The final defining characteristic of a mountain is its radar backscattering behavior at nadir. Using the altimetry data, we find that the backscatter is low, where the nadir returns are anti-correlated with the off-nadir returns (e.g., Farr, 1993; Fig. 6).

3.1.1.b. SAR-bright, Dissected Terrains (*Vd$_b$*)- **Type Example** – 85º N, 116º W - Fig. 5d/e/f

The SAR-bright, dissected terrains are abundant in both the north and south. Spatially, the unit occurs sporadically across both poles (Figs. 2b and 3b), though most often they occur at lower elevations along the borders of the largest basins. The texture appears to result from a high degree of erosional dissection, with high local slopes that give rise to the brightness in SAR. As for the mountains, we therefore assume the radar backscatter to be uniform, dominated by diffuse scattering. Initial analysis of the region suggested this was the mountain unit, yet with topographic information, the unit appears with variable elevations. As this unit does not have the highest relief, we therefore subdivide the two units to capture this effect.

In ISS, the unit appears as relatively dark patches, like the mountains, though not as dark as the lakes and seas (Turtle et al., 2009).

Both the SAR-bright, dissected terrains and the mountains were called 'dissected plateaus' and 'crenulated terrain' by Moore et al. (2014) and as 'rough highland material' by Williams et al. (2011). Lopes et al. (2010) and Malaska et al. (2016) refer to these units as hummocky/mountainous terrains and mountains respectively, which are found to be far more prevalent at equatorial latitudes (Lopes et al., 2010) and around the Xanadu region (Radebaugh et al., 2011).

### 3.1.2. Dissected Uplands

#### 3.1.2.a. SAR-Dark Dissected Uplands ($Hd_d$) - **Type Example** – 83º S, 40º W - Fig. 5j/k/l

The SAR-dark, dissected uplands are more pervasive in the south. Within any local region that does not contain mountains, the SAR-dark uplands are always the highest topographically and form drainage divides. A high density of relatively short, poorly integrated valley networks with valley widths on the order of the spatial resolution of the radar characterizes the unit. Channels in the SAR-dark, dissected uplands have variable orientations, show no dendritic or rectangular patterns (Burr et al., 2013), and they may terminate abruptly.

The relief is high for this unit, ~350 m above surrounding plains (up to 870 m), which is comparable to the relief in the mountains (*Mtn*) unit. Empty depressions (Section 3.4.1) are rarely found in proximity. At the north, this unit is most abundant in the area between the three largest seas, Kraken Mare, Ligeia Mare and Punga Mare (Fig. 2b).

In the south, this unit has greater valley widths, and follows a more organized valley network system in some locations. Similar to the north, the SAR-dark, dissected uplands are globally one of the highest topographic units, often acting as a divide between drainage basins. This unit encompasses the labyrinth terrain unit of Malaska et al. (2010; 2016) and Lopes et al. (2016).

The analysis of altimetry data shows the SAR-dark, dissected uplands to have a correlation in backscatter. At both nadir and off-nadir incidence angles, the normalized radar backscatter cross-section is characteristically low (Fig. 6b/f). This correlation in backscatter suggests that there is a material difference between the mountains and SAR-dark dissected uplands (e.g., Michaelides et al., 2016).

In ISS images, the unit appears to have a high albedo both at the north and south poles, similar to the way it appears at equatorial latitudes (e.g., Malaska et al., 2016).

#### 3.1.2.b. SAR-Bright Dissected Uplands ($Hd_b$) - **Type Example** – 75º S, 124º E - Fig. 5g/h/i

The dissected uplands may also appear bright in SAR images. This unit is the most extensive unit at the north and, like the SAR-dark, dissected uplands, is nearly always, locally, the highest topographic unit (~250 m above surrounding units). Some of the boundaries of the SAR-bright, dissected uplands also appear to be steep-sided scarps. An important distinction between this unit and mountains (*Mtn*) and the SAR-bright dissected terrains ($Vd_b$) is that the SAR-bright, dissected uplands do not always have the greatest local relief in their region, which we define as a requirement for the mountains. Furthermore, the appearance of this unit may vary between swaths of different incidence angles implying the unit's radar backscattering is less diffusely dominated than a unit like the mountains.

Numerous valleys, with narrow valley widths are found within the unit. Commonly embedded within the unit, or adjacent to it, are depressions. These depressions are often empty, and at times appear to coalesce into larger, elongated depressions. The SAR-bright, dissected uplands also segregate the north into individual basins. Previously, Lopes et al. (2010) classified this unit as "hummocky/mountainous terrain" in their global map and as "scalloped plains" in Malaska et al. (2016).

The unit associates well with a high near-IR albedo terrain that appears in ISS data as a deposit that encompasses the north (Fig. 7). It is not clear whether this unit exhibits a similar near-IR albedo in the south, as there are a large number of clouds present in the ISS observations of the south.

### 3.1.3. Mottled Plains ($Vm_b$)- **Type Example** – 82° S, 166° E - Fig. 5m/n/o

Mottled plains have a radar signature that can be highly variable (Fig. 4), which we interpret to be the result of a high degree of dissection. Topographically, this unit varies in elevation as well, with low, undulating local slopes. The mottled plains are always situated lower than the mountains and both the SAR-bright and SAR-dark, dissected uplands. Most often, these plains are in contact with the dissected uplands where they exhibit increased variability in their backscatter and gradational boundaries. Previously, this unit was mapped as "dissected mottled terrain" by Moore et al. (2014), "mottled plains" by Lopes et al. (2010), and "variable-feature plains" by Malaska et al. (2016), all of whom interpreted the unit to be eroded bedrock material.

The unit also appears in isolated patches, elevated higher than the lower lying plains units, but lower than the dissected uplands units. Associations are hard to discern in ISS observations, primarily because the scale of the unit is often quite small, and it is sporadically distributed across the polar regions.

## 3.2 UNIFORMLY SAR-DARK UNITS
### 3.2.1. SAR-Dark High Plains ($Hu_d$)- **Type Example** – 68.1° N, 128.1° E - Fig. 8a/b/c

The SAR-dark, high plains are uniformly SAR-dark and are often bounded by dissected uplands. They are situated lower than the dissected uplands and always higher than the lower plains and nearly always higher than the mottled plains. They also appear to have a large-scale slope in the direction of the seas, though they are relatively planar over smaller length scales.

The highest density of filled and empty depressions and few visible channels also characterize this unit. The SAR-dark plains are one of the largest polar units by area (Table 1). Along with the dissected uplands, the SAR-dark, high plains exhibit a relatively high albedo in ISS observations.

### 3.2.2. SAR-Dark Low Plains ($Lu_d$)- **Type Example** – 68° N, 21° E - Fig. 8d/e/f

The SAR-dark, low plains cover a large fraction of both poles (Table 1). They lie topographically lower than all other geomorphologic units except the filled depressions and seas. Stratigraphically, they are one of the youngest units, emplaced after the dissected uplands, mountains, and SAR-bright dissected terrains. Collectively they contain the greatest number of continuous valleys organized into distinct networks of any polar unit. The relief is also substantially less than the dissected uplands units. At the south, these valleys drain larger drainage basins than those in the north. At the north, all of the largest channels (Burr et al., 2013) observed on Titan are incised within this unit. One such channel, Vid Flumina (Fig. 9c) appears to have a liquid channel width of at least 150 m and a liquid elevation ~300 m below the surrounding terrain (Poggiali et al., submitted).

This unit also has a large-scale slope in the direction of the filled or empty seas. Next to the seas at the north, alluvial fans mantle this unit in some locations, though not everywhere.

In ISS images, the appearance associates most with the brightest terrains. Along with the dissected uplands ($Hd_d$ and $Hb_d$), the three units may collectively form a polar layer at the north (Fig. 7), suggesting that they may be similar compositionally.

3.2.3. Low Flat Plains ($Lf_d$)- **Type Example** – 74.8º N, 1.3º W - Fig. 8g/h/i

The low flat plains are similar in morphology to the SAR-dark, low plains, except that their appearance in SAR images is the darkest on Titan after the equatorial dunes (Lorenz et al., 2006), filled depressions, and filled seas. The unit also has minimal relief and is accordingly called 'flat' compared to the more undulating nature of the high and low SAR-dark plains ($Hu_d$ and $Lu_d$). The backscatter is also more uniform, and darker compared to the SAR-dark, low plains, and so we classify them as a separate unit.

This unit appears in the bottom of some empty depressions, which previously were classified as "granular" lakes by Hayes et al. (2008). In other instances, the unit covers large expanses of Titan's polar terrains, always topographically low, and flat over large spatial scales.

In ISS observations, an association is found with the surface features interpreted to be surficial liquids deep enough to absorb incoming photons (Brown et al. 2008). The unit is therefore different from the other SAR-dark plains units ($Hu_d$ and $Lu_d$), in that they are ISS dark, rather than bright (Malaska et al., 2016).

### 3.3 UNIFORMLY SAR-BRIGHT UNITS

3.3.1. Uniform SAR-Bright Plains ($Vu_b$)- **Type Example** – 64.7º N, 22.7º E - Fig. 10

This terrain unit is distinctively SAR-bright, with a uniform radar signature at our mapping resolution. The uniform SAR-bright plains contain fewer filled lake depressions than the SAR-dark, high plains ($Hu_d$), while they also lack any discernible valley networks. When in contact with the dissected uplands, the uniform SAR-bright plains appear more variable in both elevation and brightness. In these scenarios, they were mapped as the mottled plains ($Vm_b$) instead. The unit also differs from the dissected uplands in both their reduced relief and lack of observable dissection.

The uniform SAR-bright plains are more abundant in the south than the north, and in most cases, they are located far away from the large filled/empty sea structures that dominate both poles. Some of the largest alluvial fans also deposit over this unit in the south. We are unable to determine any spectral characteristics of the unit in ISS data because of their small spatial extents.

### 3.4 SUPERPOSED UNITS

3.4.1. Filled/Empty Depressions ($Fl/El$) - **Type Example** – 68º N, 130º E - Fig. 9a

Depressions appear as both liquid filled and empty. We refrain from using the nomenclature of "granular" or "partially-filled" lakes (Hayes et al., 2008) as recent altimetry results suggest that these depressions may be filled by many meters of liquid if similar in composition to Ligeia Mare (Mastrogiuseppe et al., 2014), and still be undetectable in SAR images. The depressions are clustered into distinct regions at both poles (Figs. 2b and 3b). Often appearing together, they are embedded up to ~600 meters deep (Hayes et al., submitted) into the SAR-dark, high plains unit (Figs. 2b and 3b). Some of the depressions also appear to have raised rims that can extend hundreds of meters higher than the surrounding terrain (Michaelides et al., 2016; Hayes et al., submitted).

Empty depressions become more prevalent where these depressions are in drainage basins that drain into the large seas. In these locations, the elevations of the floors of the empty depressions are always higher than the elevation of the sea level of Titan's three seas, suggesting a hydraulic connectivity over these relatively short length scales (Hayes et al., submitted). When these depressions are proximal to the seas, their boundaries become more irregular in planform.

Liquid-filled depressions further from the seas appear darker in SAR images than those nearer to the seas. Recent observations have demonstrated that the methane-rich liquid in Ligeia Mare is very transparent at centimeters wavelengths (Mastrogiuseppe et al., 2014; Mitchell et al., 2015). This implies that the liquids in these depressions must be hundreds of meters deep if they are similar in composition to Ligeia Mare. A more plausible alternative that can be tested using Cassini's altimeter, is that these depressions contain liquids that are considerably less methane rich than Ligeia Mare (Mitchell et al., 2015; Hayes, 2016). Empty depressions nearer to the seas are typically lower in elevation and display darker floors than those located further from the seas (Hayes et al., submitted).

In the south, these depressions also form in clusters; and the shape of boundaries suggest that separate depressions formed and then coalesced into larger structures than those at the north (Fig. 11a/b). The total area of filled depressions is less in the south than in the north, though the area of empty depressions is greater at the south (Table 1).

A minority of the depressions may also have very circular boundaries. Though still depressions, they appear different from the other depressions, in that they are more circular, and at times have a bright, mounded rim around their boundaries (Fig. 9b). They are often observed to form in clusters at the boundaries of, and also visible under the liquid surface level of, some of the larger depressions and seas that are more transparent to the radar (Fig. 12c-iii). Additionally, these depressions appear to have a characteristic diameter of ~5–10 km, smaller than the vast majority of other depressions on Titan. Their shape, distribution, and size may provide insight to their formation mechanism.

Previous interpretations for the origin for some these features included cryovolcanic calderas (Wood et al., 2007), karst sinkholes (Mitchell et al., 2007) and impact craters (Wood et al., 2010).

*3.4.2* Filled/Empty Seas (*Fm/Em*) - **Type Example** – 79º N, 107º E - Fig. 13

The seas that cover a large fraction of the mapped north polar region (e.g., Hayes et al., 2008: Table 1) are included here as an embedded unit. As noted previously, the SAR-bright, dissected uplands unit compartmentalizes the north into individual basins. More highly dissected SAR-dark, dissected uplands segregate the south into similar basins. Basins that are filled with liquid are Punga Mare, Ligeia Mare and Kraken Mare at the north. Ligeia Mare (Fig. 13b) was measured directly by the altimeter to be ~160 m deep (Mastrogiuseppe et al., 2014, 2016).

The south is dominated by four large basins, interpreted to be empty seas (Wood et al., 2013; Hayes 2016), and a similar cluster of empty depressions as the north. The total sea areas at both poles are similar, with the south ($7.6 \times 10^5$ km$^2$) slightly larger than the north ($7.0 \times 10^5$ km$^2$). Presently, the empty seas at the south, however, do not appear as interconnected at the surface as the flooded northern seas (Hayes et al., 2011). The interpreted shorelines of these empty seas show a similar morphology to the filled seas in the north, with the largest channel inlets and paleo-island structures appearing around their perimeters. A SAR-dark plains material floors these empty seas (e.g., Fig. 13a). Ontario Lacus, the largest liquid-filled depression in the south (Brown et al., 2008; Wall et al., 2010) is located at the lowest point of one of the empty seas (Wood et al., 2013).

In ISS images, the empty seas are dark, similar in appearance to the SAR-dark, low plains. This pattern is different from the undifferentiated plains (Lopes et al., 2016; Malaska et al., 2016) around the mid-latitudes and equatorial regions of Titan, which are similar in SAR appearances though distinctly bright in ISS observations (Malaska et al., 2016; Lopes et al., 2016).

3.4.3. Fluvial Valleys - **Type Example** – 73º N, 117º E – Fig. 9c

Numerous valley networks dissect Titan's polar terrains (Jaumann et al., 2008; Malaska et al., 2011; Langhans et al., 2012; Burr et al., 2013; Black et al., 2013). As individual channels are not observed in SAR, we refer to their mappable portions as fluvial valleys. The valleys in the dissected uplands units are narrow and are embedded in terrains of high relief. As the cumulative drainage area increases downslope towards the sea, the valleys widen, which on Earth would be consistent with an increasing fluid discharge downstream. Valleys near the seas, where there is altimetry data available (Vid Flumina: Fig. 9c), are incised ~300 m into the surrounding terrain (Poggiali et al., submitted) suggestive of a prolonged period of incision and sediment transport across the region.

In the south, there are a greater number of valleys and a higher fraction of those valleys are SAR-bright, like many of the equatorial valleys around Xanadu (Le Gall et al., 2010).

Valleys distant from the seas that are not inundated with fluids are difficult to analyze with ISS images because of the small spatial scale of these features. The valleys closer to the seas have a similar ISS signature as the seas themselves.

3.4.4. Alluvial Fans (*Af*) - **Type Example** – 68º N, 76º E - Fig. 9d

Alluvial fans are distinguished by their fan-shaped, relatively SAR-bright, undissected surfaces. They are located at the termini of valleys crossing the boundaries of the mountains and SAR-bright dissected terrains units. Alluvial fans are rarely found around the dissected uplands. These features are also contained within the mottled plains ($Vm_b$), suggesting that the mottled plains may be a fluvially emplaced transitional unit.

In some areas, fans overlap to form a bajada. This unit acts as a morphologic indicator as to the sources, sinks, and transport paths of sediment on Titan (Radebaugh et al., 2016; Birch et al., 2016). We mapped 28 fans in total, with the majority (21) residing in the north. The fans in our mapping regions are too small to resolve in ISS observations.

# 4. GEOMORPHOLOGIC UNITS: INTERPRETATIONS

By incorporating topographic data as an inherent characteristic in the definition of our geomorphologic units, we are able to generate a schematic illustrating the relative topographic relations of our units (Fig. 14). In Fig. 14, units that are topographically the highest are at the top of the column. For example if the SAR-dark, high plains ($Hu_d$) are found next to SAR-Bright Dissected Terrains ($Vd_b$), the dissected terrains will always be topographically higher. Stratigraphically, however, the plains would overly the dissected terrains. This column allows us to compare elevations and contact relations, which aid in the development of our model in Section 5.

The same geomorphologic units can be mapped at both of Titan's poles, suggesting that the governing processes shaping these landforms have been the same. While the south is currently lacking large inventories of exposed surface liquids, there are still large empty seas, with SAR-dark floors and numerous empty depressions.

*4.1. Mountains & SAR-Bright, Dissected Terrains ($Vd_b$ & Mtn):*

We classify the SAR-Bright Dissected Terrains ($Vd_b$) and Mountains (*Mtn*) as distinct units in our mapping because of their differences in topographic relief. Yet the similarity in morphology, appearance in ISS images, and altimetry backscattering suggest a similar formation and water ice-rich composition for these two units. Across the poles, we see SAR-dark plains units at higher elevations than the SAR-Bright Dissected Terrains, which suggests a mantling of the plains units on top of the dissected terrains. In many locations, it appears that the SAR-Bright Dissected Terrains are outcropping from the plains units, appearing as exposed regions of the underlying bedrock. Most clearly, these units (hereafter both *Mtn* and $Vd_b$ are termed mountainous terrains) occur around the western perimeter of the northern seas and around the empty southern seas. In such cases, these regions may represent areas that have had a sedimentary cover removed, similar to the degraded craters observed at Titan's equatorial region (Soderblom et al., 2010; Neish et al., 2015).

Similar to Lopes et al. (2010), we interpret the two units together to comprise the primordial crust of Titan, acting as the oldest terrain unit in the north. Where the terrains occur at the highest elevations, they are mapped as mountains. Elsewhere, where there may be adjacent units that are topographically higher, they are mapped as SAR-Bright Dissected Terrains.

The quasi-linear orientations of the mountainous terrains at the south may imply that their formation and evolution may be tied to tectonic processes, such as polar subsidence (Choukroun and Sotin 2012; Moore et al., 2014), true polar wander (Schenk et al., 2008), primordial remnants as Titan despun (Cook-Hallett et al., 2015), or some combination of endogenic process (e.g., Liu et al., 2016). It is difficult to determine any preferred orientation of mountainous terrains in the north.

*4.2. Dissected Uplands ($Hd_d$ & $Hd_b$):*

We separate the dissected uplands from the mountainous terrains because our observations suggest that they differ in composition, even though they also exhibit high relief. Compared to the mountains, the dissected uplands are less bright, in SAR lacking very clear layover texture (e.g., Lopes et al., 2010; Malaska et al., 2016). These scarp-bounded units are highly dissected, with a higher channel density than the lower lying plains. The SAR-bright dissected terrains are topographically emplaced lower than the mountains, though stratigraphically they are situated above the mountains.

The defining characteristic of the dissected uplands, however, is their backscattering behavior at nadir. Using the altimetry datasets, we find that the scattering is not dominated as diffusely scattering, where the returns are correlated between nadir and off-nadir incidence angles (Fig. 6). For the mountainous terrains, this is not always true (Fig. 6). This indicates that the high backscatter that we observe in SAR images for the dissected uplands is not as diffusely dominated, but instead may be the result of a different composition or subsurface structure that causes absorption at all incidence angles. The dissected uplands have similar ISS signatures to the SAR-dark, high plains (Fig. 7), and so we hypothesize that the dissected uplands terrains may be composed of more organic products than the more water-ice rich mountainous terrains. Malaska et al. (2010) interpreted the SAR-dark, dissected uplands as karst terrains resulting from dissolution processes while Malaska et al. (2016) and Janssen et al. (2016) identified the radar characteristics as consistent with organic materials.

The dissected uplands in the north, where mountains are absent, also act as drainage divides between basins. Furthermore, the drainage basins that these units enclose often do not empty into the seas. Instead, they are distant from the seas, forming what appear to be closed basins, and within these close basins, the SAR-dark, high plains are the dominant structure, themselves cut into by filled and empty depressions. A similar type of pattern seems to occur at the south, also clustered, elevated, and distant from the large empty seas.

These drainage basins are also topographically situated higher than the filled/empty sea basins. The density of filled and empty depressions is highest in these drainage basins, with the density the highest in the north polar basins. Very few channels are observed in these basins. That does not imply, however, that there are no channels (or other hydrological features) in the region, as numerous channels may exist at finer scales (less than ~150 m). Future high-resolution altimetry profiles may be able to detect such features.

We forward two possible ways in which this change in composition might have occurred. In the first scenario, the dissected uplands would form as a primary deposit due to the evolution of the hydrocarbons in the desiccating seas that left a layer of less-soluble hydrocarbons as the seas retreated. In the second scenario, the dissected uplands would be a residual deposit of insoluble hydrocarbons that covers upland areas and resists erosion. This would imply that they are younger, resistive sediments that form a 'cap-rock,' which would explain why they are found at higher elevations and along the drainage divides.

*4.3. SAR-Dark, High Plains ($Hu_d$):*

The SAR-dark, high plains are commonly found within the elevated drainage basins that are characterized by the filled and empty depressions. We interpret this unit accordingly as the dominant depression-forming unit (Fig. 11c/d). The SAR-dark appearance of this plains unit may result from a saturated top surface that absorbs incoming radiation. The SAR-dark appearance may also be the response from a flat, fine-grained substrate like the equatorial dunes (Lorenz and Radebaugh, 2009), suggesting these plains may be a part of a larger sedimentary deposit of clastic material. If there are no sub-resolution channels, the unit is then likely to be moderately permeable, allowing subsurface fluids to flow great distances (Hayes et al., submitted). We observe this relation in the region south of Ligeia Mare, where many empty depressions occupy the higher standing terrain, while downslope the depressions appeared to be liquid filled (Fig. 11c/d). This observation suggests that at higher liquid levels, the empty depressions may fill as well.

The formation of depressions within this unit suggests that its composition and material properties are conducive to the formation of such features. Further, the depth of the depressions (up to ~600 m; Hayes et al., submitted) suggests that the layer must also extend to such depths in those locations. The mountains and SAR-bright dissected terrains also appear to outcrop from this unit, suggesting that the SAR-dark, high plains are a mantling layer. As such, we interpret these plains as sedimentary deposits that have been emplaced on top of the underlying bedrock.

*4.4. Uniform SAR-Bright Plains ($Vu_b$):*

The uniform SAR-bright plains have a similar backscattering appearance to the floors of the empty depressions, though the plains lack any boundary that would distinguish them as enclosed basins. Instead they appear as broad plains. We interpret their origin to be similar to the lower, SAR-dark plains units ($Lf_d$ and $Lu_d$). The difference in backscatter may then be attributed to larger, centimeter-sized grain sizes on the surface of the brighter, higher plains. In such cases these plains could be analogous to gravel piles with cm-sized scatterers distributed across their surfaces. The increased brightness of the uniform, SAR-bright plains may also be due to an increase in volume scattering, where the uniform SAR-bright plains have greater scattering in the subsurface. This could be the result of a more porous subsurface and/or a subsurface that lacks fluid within the top few centimeters (Janssen et al., 2016).

*4.5. Mottled Plains ($Vm_d$):*

The mottled plains have a highly variable appearance in SAR images. They also appear at variable elevations across the poles, though always lower than mountainous terrains, dissected uplands, and uniform SAR-dark high plains. This suggests that they are a transitional unit between the higher units and the lower-lying SAR-dark plains ($Lf_d$ and $Lu_d$). We attribute the variable backscattering and elevation of this unit to be the result of varying degrees of fluvial incision and deposition at the boundaries of the topographically high regions. Further, the association with alluvial fans suggests a common origin that would be the result of fluvial incision and varying transport capacities between units of varying elevations.

The mottled plains also appear in isolated patches, elevated higher than the lower lying plains units. This may represent an outcropping of an underlying substrate, and in such a case, may be similar to the hummocky sections of the hummocky/mountainous terrains identified by Lopes et al. (2010).

*4.6. Low Flat Plains ($Lf_d$):*

The north is also composed of relatively low relief areas which contain the largest filled depressions (Bolsena Lacus; Fig. 2b). These regions are especially complex, as the low flat plains cover a large fraction of these areas. The flatness and uniform darkness in SAR of these regions, over large spatial scales, can be interpreted as evidence that the phreatic surface may be near to the surface, possibly ponding in the lowest areas and/or at the bottom of some empty depressions. Alternatively, the low, flat plains may also be regions of radio wave-absorbing, sedimentary material that are uniform and flat; similar, though distinctly different in origin, to the equatorial dunes (e.g., Lorenz et al., 2006). We favor the former, because of the association of this unit with ISS regions interpreted to be liquids.

*4.7. SAR-Dark, Low Plains ($Lu_d$):*

In regions closer to the mare, the SAR-dark, low plains are emplaced lower than the dissected uplands and mountains terrains. The inclined surfaces of this unit and high density of observable fluvial valleys suggest that it must postdate the formation of the seas and be formed by surface processes, most likely being rivers and alluvial fans, that drain towards the seas. We therefore interpret this unit as sedimentary alluvium transported from the high relief regions onto the adjacent lowlands. Larger water-ice sediment transported as bedload (e.g., Perron et al., 2006) and materials carried in suspension would tend to deposit where the valleys crossed the boundaries to the low gradient adjacent units where reduced slope would decrease the flow boundary shear stress (e.g., Leeder, 2011).

Within this unit, the 5-μm bright unit identified by Mackenzie et al. (2014) is frequently found, perhaps suggesting a causal relationship. If so, their interpreted evaporitic nature (Barnes et al., 2011) would just be one limiting case, where runoff from the dissected terrains is temporarily ponded within empty depressions and allowed to evaporate.

Evidence for a substantial, compositionally heterogeneous, sedimentary layer at the poles is also seen, where ~300 m deep canyons are incised within the SAR-dark, low plains unit. Changes in erosional resistance help to explain these features, where a soft layer may be overlain by a harder cap rock. Once the river has eroded through the resistive cap rock it would be able to incise deeply into underlying softer layers. An analogous stratigraphy may also explain the morphology of the depressions, particularly those with raised rims (Hayes et al., submitted).

While the SAR-dark, high plains, and the SAR-dark, low plains are morphologically similar, we classify them separately on the basis of topographic emplacement. If the units share a similar origin, as in the case of the mountains and SAR-bright dissected terrains units, then we are observing sedimentary layers that form a stepped topography with variable elevations across the poles. If the units were emplaced as one contiguous layer, then the large-scale slope of both units and stepped topography would suggest that vertical crustal motions displaced these units to their current positions. Detailed spectroscopic analysis should be able to test this hypothesis and yield further insight as to whether they share a common origin.

*4.8. Filled/Empty Depressions (Fl/El):*

Our mapping of the filled and empty depressions at the poles completes that of Hayes et al. (2008). We notice, however, that not all filled depressions have similar appearances in SAR. Filled depressions are relatively darker in their appearance when they are in drainage basins distant from the seas as mentioned in Section 3.4.1.

# 5. EVOLUTIONARY MODEL

Despite the difference in the distribution of filled depressions and seas, Titan's north and south polar regions have similar morphologies. Accordingly, we propose that the processes that formed their surfaces were similar (e.g., Dietrich et al., 2003). The major differences between the poles are the greater density of mountains and SAR-bright dissected terrains in the south and the greater areal coverage of liquids in the north. In the model we discuss below, we adopt the concept of Aharonson et al. (2009), where fluids are dominantly transported from pole-to-pole over the ~100,000-year apsidal precession of Titan's orbit (Lora et al., 2015).

The landscape of Titan, however, is complex and is probably the result of surface processes acting over many millions of years. Any model for the geomorphologic evolution of Titan has to explain large variations in radar backscatter and topography, as well as a variety of landforms including: rivers (Burr et al., 2006), canyons (Poggiali et al., submitted), lakes and seas, and mountains (Radebaugh et al., 2007). Accounting for the formation of the filled/empty depressions is also fundamental to any model, where their closed, steep sided nature must imply either dissolution or sublimation processes acting on a saturated, porous substrate (Hayes et al., submitted). Further, the raised rims of the largest empty depressions, which are characteristic of the largest of these features (Hayes et al., submitted), suggest an added complexity that has no applicable terrestrial analog.

The presence of sloping geomorphologic units ($Hu_d$ and $Lu_d$) is evidence for the development of relief within the drainage basins, either through uplift of the surrounding mountains or subsidence and erosion. Further, simply dissecting the ocean perimeters would not leave the large basins that the seas occupy. Producing the current landscape without some form of surface deformation to create relief would be more difficult to conceive, and so we assume that endogenic, relief-generating processes have occurred.

Detailed modeling (e.g., Tewelde et al., 2013; Neish et al., 2016; Howard et al., 2016), though not within the scope of this work, may yield results that match the form of the landscape we have found in our mapping. We explore one possible evolutionary scenario that is able to describe the bulk of the distribution of the mapped units, while acknowledging that other models may equally describe the landscape our mapping reveals.

*STAGE 1 – Build-up of Large Sedimentary Deposits & A Polar Ocean (Fig. 15a)*

The model illustrated in Figs. 15 and 16, assumes that in Titan's distant past the polar regions were submerged beneath a methane-ethane ocean (e.g., Lunine, 1993) of sufficient depth and for sufficient duration that thick sedimentary layers accumulated on their seafloors. An ocean would be the product of episodic outgassing of methane driven either by endogenic processes like cryovolcanism (Lopes et al., 2007), through the liberation of methane from clathrates in Titan's crust (Tobie et al., 2006), or from the late volatilization of nitrogen and methane surface ices as the sun warmed (Moore et al., 2014). The later options would allow for multiple ocean epochs and is more favorable given the scarcity of confirmed cryovolcanic features on Titan's surface (Lopes et al., 2013).

We propose that the topography of the ancient water-ice landscape (marked by the top of the red unit in Fig. 15) defines local highs and lows that may have developed in response to erosion of large impacts structures or uplifted terrain due tectonic processes. The low areas then become sites of ocean basins. The sediments that accumulated on the seafloor of these basins are then presumed to contain components of both organic atmospheric products and water-ice sediment that was eroded from the surrounding landscape. A terrain composed of both water-ice sediments and organic components was also postulated by Neish et al. (2015) to explain the spectral character of impact craters. In this model, the mountainous terrains (*Mtn* and $Vd_b$) represent regions where this primordial bedrock remains exposed, either never having been submerged below an ocean or recently exhumed.

Some organic products are soluble in liquid methane or ethane, such as benzene, naphthalene, and biphenyl (Malaska and Hodyss, 2014). These materials will remain in solution, forming evaporitic deposits around lakes and seas that have lost liquid due to evaporation. Many of the other organic compounds are not soluble (e.g., tholins, poly-HCN; Raulin et al., 1987;

Cornet et al., 2015) and will instead form a layer, or layers, of sediment on the ocean floor and/or a lag deposit on uplifted, exposed surfaces.

Accordingly, we would expect these sediments to form layered deposits with varying compositions and strength properties between soluble organic materials, insoluble organic materials, and insoluble water-ice sediments (Cordier et al., 2013; Cordier et al., 2016). Further, these layered deposits, having variable compositions, solubilities, and erodibilities, are likely to weather and erode differently. This would allow the landscape to develop complex topography and relief. As the methane and ethane ocean evolved over time, potentially forced by orbital cycles, the composition of the sedimentary layers deposited within the ocean may have changed as well, adding a further complexity to the landscape's evolution.

The presence of a polar ocean is important for two reasons. First, an ocean concentrates the sedimentary deposits ($Hu_d$) from the atmosphere and surrounding terrains into discrete regions at the poles. Photolytically generated products are globally deposited, but more volatile species are primarily transported through the atmosphere to the cold-trap at the poles (Brown et al., 2006). If the polar regions are preexisting global topographic lows, significantly lower than the present day (Lorenz et al., 2013), then over geologic time scales the organic products will also be transported to the poles as clastic sediments, and deposited at the bottom of the proposed polar ocean. Water-ice sediments, also transported as clasts, would be deposited within these ocean basins as well.

The ocean would also act as a protective cover, allowing for the buildup of vertically thick sedimentary deposits before the onset of various erosive processes that act on the sediments themselves following aerial exposure. The complete lack of confirmed impact craters at the poles further suggests that these regions may have been covered by liquid bodies for significant periods of Titan's history (Neish and Lorenz, 2014).

*STAGE 2 – Exposure of Sedimentary Deposits & Formation of Depressions (Fig. 15b)*

Over geologic time-scales, we assume that the ocean retreats to higher latitudes as methane is photochemically dissociated in the upper atmosphere and irreversibly converted into more complex compounds (e.g., Larsson and McKay, 2013; Moore et al., 2014). As the ocean retreats, a large inventory of precipitable methane (in the atmosphere) could still be present, where rainout may still cause intense erosion of exposed sediments, except in the lowest portions of the poles that remain liquid filled. In Titan's past, events as described may have also happened multiple times as methane is episodically outgassed and photochemically destroyed (Tobie et al., 2006). In such case, the cycle would restart at the earliest stages.

Throughout this second evolutionary stage, the seafloor below the shallowest parts of the paleo-ocean gradually become more exposed, where they are subsequently eroded, revealing the water-ice-rich mountain basement beneath. Erosion is assumed to progress at differential rates, depending on the composition and erosional resistance of the exposed layers, resulting in the stepped topography (illustrated by a graben in Fig. 15). The relative importance of mechanical versus chemical (dissolution) erosion likely varied, depending on what sedimentary materials are exposed, at what elevations they appear in the landscape, and for how long they are exposed. Such complex layering may explain some of the observed geomorphology, for example, the deep canyons and filled/empty depressions, which suggest a resistive cap rock overlaying a soft layer.

Because both the SAR-bright, dissected uplands ($Hd_b$) and the SAR-dark, high plains ($Hu_d$) contain the majority of the filled/empty depressions, their compositions are likely similar, though not identical, and we suggest that they may share a common origin. The SAR-dark, dissected uplands ($Hd_d$), on the contrary, appear less conducive to the formation of filled/empty depressions, suggesting they differ in composition from these other units.

We assume that atmospheric deposition of organic sediments has continued throughout all stages of our model, at rates proportional to both the available amount of methane and high-energy photon flux from the Sun. The concentration of the volatile materials is also presumed to be at the poles (Brown et al., 2006). As the ocean retreated, much of the methane inventory would be periodically transported from pole-to-pole as in the model described by Lora et al. (2015). When the ocean was deep enough, this cyclic exchange of fluids would be most felt along the shorelines, where significant loss of fluid (while maintaining a sufficient precipitable volume) will cause sea-level to drop, and a pulse of channel incision to propagate away from the shorelines. Evidence of this is seen in the shorelines of the southern empty seas (e.g., Hayes, 2016) with perimeters composed primarily of the mottled plains ($Vc_b$). At the opposite pole, there is a change in the fluid balance that is causing liquid levels to rise (currently the north). The lack of any observable deltas along the shorelines of the northern seas implies that total amount of landscape erosion and/or sediment transport has not been sufficient to support the formation of large deltas to develop into the seas. This apparent paradox may be explained either relatively recent rise is sea level or by the delivery of fluid to the seas that in a manner that didn't cause significant erosion of surrounding landscapes.

Throughout the retreat, as more sedimentary materials become exposed, the depressions would then begin to form. Evidence for the formation of small circular depressions in substrates initially deposited at the bottom of liquid bodies is seen across the north. Beneath the liquid surfaces of both Jingpo Lacus (Fig. 12c-iii) and Ligeia Mare, and around the shorelines of all large bodies of liquid (Kraken, Punga and Ligeia Mare, Jingpo and Bolsena Lacus) small circular depressions (Section 3.4.1.b) are found, suggesting that such substrates promote the formation of depressions.

As such, we hypothesize that the seas are the ultimate sinks for polar sedimentary materials, and that their substrates are similar in composition to the SAR-dark, high plains ($Hu_d$). Initially there would be more filled depressions, but as surface liquids are lost, and the phreatic surface lowers, empty depressions become more prevalent. Further, the continuous variation of base level over geologic time, both in varying liquid elevations due to orbital variations and, over longer timescales, tectonic uplift/down-dropping (vertical dashed lines feature in Fig. 15), would act as a major control on the formation of the landscape across the polar regions. The impact of these variations likely ties into the formation of the depressions as well. Because depressions are not found at the bottom of the empty seas, we suggest that their exposure for extended periods must be relatively recent. If their formation is dependent on subsurface flow, then an accompanying high alkanifer level during an unexposed time period can slow and/or delay their formation to subsequent periods when liquid levels are lower. Alternately, the lack of depressions around and within the southern empty seas may be explained if there remains an insoluble lag that is preventing the initial growth and formation of these features.

The variation in both surface and subsurface liquid elevations will force the landscape to continuously readjust itself, where each exchange of fluids slowly acts to form the filled/empty depressions and sea shorelines that we see today.

*STAGE 3 – Current Day: Ocean Retreat Reaches Critical Level (Fig. 15c)*
Methane is assumed to be progressively lost over time (e.g., Larsson and McKay, 2013; Moore et al., 2014), and eventually the seas in the evacuated poles would become nearly entirely dry when fluids move to the opposite pole. Such a scenario would expose the former seafloor, and fluids would remain in only the most pole-ward locations. Over time, the size and number of remaining liquid bodies will decrease, eventually reaching the current state of only having a very small number of filled liquid depressions in the evacuated pole.

Empty depressions would then become increasingly common, where depressions at higher elevations would be abandoned as they would no longer be connected to an aquifer. With higher resolution topographic information, we would expect to find the boundaries of these depressions to be more highly degraded relative to the presumably younger depressions at lower elevations, closer to the seas. In locations where empty depressions are found adjacent to filled depressions (Fig. 11c/d), the empty depressions are located at slightly higher elevations (Hayes et al., submitted), and in such cases they too may be abandoned, presumably having been filled during a previous epoch with larger surface liquid inventories.

Without any subsequent methane outbursts to restart the cycle, over time the filled and empty depressions will erode and eventually coalesce into larger depressions. Such a scenario seems to have progressed at the south already (e.g., Fig. 11a/b). The growth of these features may reach the point where enough depressions have merged and a single, enclosing perimeter will no longer be noticeable. The remaining boundaries would be highly degraded, with large slumps and lower slopes than a fresher depression. The empty depression in Fig. 11a appears to have resulted from the progression and growth of multiple depressions, where the leftward boundary in Fig. 11a (red arrow) is no longer visible. Further, the degradation of depressions at the evacuated pole is likely accelerated if the depressions are in communication with the seas, as they are in the north (Hayes et al., submitted). The most likely locations for a more rapid degradation of depressions would be around the perimeter of the current sea shorelines, where the local phreatic surface may be closer to the surface. The depression in Fig. 11a is also near the border of an empty sea, providing evidence that such a situation could be occurring.

In locations where there are local topographic lows, sediment can collect and lead to the creation of the SAR-dark low plains ($Lu_d$). These dark plains materials also seem to border highland regions around the seas, suggesting that they are sedimentary materials transported and deposited at the base of the uplands. Where enclosed regions of this unit intersect the local phreatic surface, liquids may be able to pond at or near the surface, creating the SAR-dark, low flat plains ($Lf_d$).

We also see many more filled/empty depressions in drainage basins dominated by the SAR-dark, high plains ($Hu_d$). If the growth of these depressions is relatively rapid over geological timescales, then without a mechanism to rebuild the sedimentary deposits into which they are embedded, we should not see any of these depressions. Because we do see such depressions, the growth of depressions must be geologically slow.

Alternatively, the depressions may have formed in the geologically recent past and we are observing Titan during a special time, or there may be some unknown process that slows the rate of growth as the depression gets bigger. Without any knowledge of the composition and elevation of liquid(s) contained in the filled depressions, and the composition and elevation of the surrounding terrains, there remains no mechanistic model that can describe the formation and evolution of these features (Hayes et al., submitted). As such, all scenarios mentioned remain possible.

In the south, the sedimentary deposits associated with the proposed ocean are less extensive, with more of the primordial underlying crust exposed as mountains ($Mtn$) or bright, dissected plains ($Vd_b$). This situation may have resulted for one of three reasons: (1) the ocean was less extensive at the south, depositing a shallower cap layer; (2) erosion rates were more effective, removing sedimentary materials at a different rate; or (3) increased tectonic uplift or subsidence of basins generated greater relief at the south, enhancing river gradients and erosion. We note that all three may contribute to the observed differences between the poles.

# 6. MODEL IMPLICATIONS & ANALOGS

The adoption of a retreating polar ocean model was chosen primarily because of its simplicity and ability to explain the observed geomorphology of the polar terrains. Our inference of dissolution on the spatial scales that we observe on Titan requires a substantial buildup and then removal of material. The presence of an ocean allows these processes to happen independently and sequentially. An ocean is also consistent with both the entrapment of volatiles in the polar regions, and a source for the global hydrological cycle.

A terrestrial analog for the model we propose can be found with the Mediterranean Sea, which has thick evaporite layer at its base. This layer formed in the late Miocene, when the Mediterranean Sea dried up during the collision of the African and European plates (Hsu et al., 1973; Ryan, 2008, 2009). Left behind was a halite and gypsum layer kilometers thick, containing ~$10^6$ km$^3$ of evaporitic material. Within the evaporites, there are repeated cycles of layers that formed as a result of the precipitation of minerals with different solubilities. As the sea level dropped, not only were these evaporitic materials deposited in large quantities, but inflowing rivers also incised and propagated away from the basin perimeter. One such example is the Nile River canyon, which incised ~570 m below current sea level into the surrounding terrain (Hsu et al., 1973; Woodward et al., 2007).

In the case of Titan, we propose that the large SAR-dark, high plains to be underlain by sedimentary deposits, analogous to the large salt layers of the Mediterranean Sea, where it is likely that there will be cycles of layering of organic molecules of different solubilities. However, the thickness of pure organic evaporite will be extremely thin on Titan unless it is constantly recharged by precipitation from the atmosphere. Evidence of past channel incision and canyon formation is also found in the channels surrounding Ligeia Mare, only that in the current epoch these valleys are drowned by rising liquid levels at the north (Hayes et al., 2011).

Larsson and Mackay (2013) began modeling the extent of an ocean on Titan, given the current rate of methane loss, and found that an ocean would extend to equatorial latitudes 600 Ma. Further, observations of the oldest terrains on Titan are consistently found at the highest elevations, including craters (Neish and Lorenz, 2014) and mountain chains (Liu et al., 2016). The presence of large liquid bodies at the poles within the last few hundred million years could also help to explain why we do not see any confirmed craters at the poles, since impacts into marine environments have muted, if any, topographic expression (Neish and Lorenz, 2014). Though we are unable to constrain the total areal extent of any past polar ocean, our model requires only the presence of larger liquid bodies at the poles. Higher resolution and a greater coverage of topographic data, which is currently not available, should be able to constrain the locations and depths of any past ocean. The acquisition of such data would require a future mission to Titan that is able to resolve features such as shorelines that follow an equipotential (e.g., Perron et al., 2007). An external constraint for both the depth and latitudinal reach of any putative ocean though, is Titan's non-zero eccentricity and limited degree of tidal damping (Sagan and Dermott, 1982).

The initial deep depression of the poles is critical to the model. This makes the poles substantially different from the terrains in the mid-latitudes or equatorial regions (Lopes et al., 2010; Malaska et al., 2016; Lopes et al., 2016). The poles would have to have been depressed a sufficient amount to prevent any paleo-ocean from spreading to lower latitudes. Depending on the rate and location of vertical crustal movement, however, the initial thickness of any ocean(s) will vary. Greater vertical motions require a smaller ocean, while reduced tectonics and uplift requires larger, deeper liquid bodies. The current ~1 km difference between the poles and equator (Zebker et al., 2009; Lorenz et al., 2013; Mitri et al., 2014) may also only be a lower limit, as significant sedimentation and infilling of the initial depression could mask a deeper initial depression. Further, gravity measurements of Titan suggest that a significant removal of material and/or deposition of sediments are required globally to explain the current geoid-topography relationships (Hemingway et al., 2013), if the outer ice crust is assumed to not be isostatically compensated (Mitri et al., 2014; Lefevre et al., 2014). With a large deposit of sedimentary material at the poles, our model aligns with the gravity model of Hemingway et al. (2013).

Titan's channel networks also suggest that the polar landscapes are not in equilibrium with their environment. We observe disorganized and drowned channel networks, and only rarely do we observe deltas. The later suggests that sediment delivery is slow relative to the rate of backwater drowning of valley networks. This adds credence to our model, in that the model relies on Titan's atmospheric and tectonic environments to be continually varying over both long (retreat and loss of methane), intermediate (uplift and down-warping of Titan's crust), and shorter (~100,000-year Milankovitch cycles) timescales.

The greater abundance of mountains in the south, suggests that the sedimentary layer of the high dark plains is not as prevalent or effective at mantling the underlying substrate there, or that the south polar mountainous terrains are topographically higher than the buried north polar mountainous terrains. The lack of a cover at the south, however, does not require different processes acting between the two poles. The same erosional processes are likely doing work on the surface through geological time, with only the occurrence rates of erosional events differing.

The lowest topographic units at either pole are the filled/empty seas. If the southern empty seas are former liquid-filled basins, then they may be expected to contain evaporitic, 5 μm-bright material at their bases, something that is not observed (Mackenzie et al., 2014). Mackenzie et al. (2014) offer two explanations to explain their absence: evaporite formed in the south but were subsequently removed or covered, or that conditions were never suitable for evaporite formation. Our model is more consistent with the evaporitic deposits having been formed and then buried, removed or altered, as there are likely many processes able to do so over 100,000 year orbital cycles (Lora et al., 2015). Our model is less consistent with a scenario in which the composition of the ocean, or the availability of soluble material varied significantly between the poles, or that the liquid did not evaporite from the south, but instead was drained away, carrying the dissolved organic material with it.

## 7. DISCUSSION & CONCLUSIONS

Our mapping reveals a complex terrain of both underlying mountainous topography and smooth undulating plains materials into which all of the lacustrine depressions are embedded. The landscapes of the south and north poles of Titan are similar, with the south pole appearing only as a drier, more eroded version of the north pole, and lacking more of the sedimentary units. Most of Titan's filled and empty lake depressions are located in regions underlain by the geomorphological unit, suggesting that this unit is conducive to the formation of these features. Dissected uplands units, compositionally distinct from underlying mountainous bedrock, border these lake-forming regions, forming enclosed (endorheic) basins.

The polar ocean model we propose is our best interpretation of the mapping. Testing of the conceptual model could be accomplished by modeling ocean-atmosphere interactions to explore if sizes and duration of an ocean are compatible with topographic data and geomorphic mapping. Tidal modeling of a Titan ocean may similarly constrain the maximum size and depth of an ocean, and define under what conditions these water bodies would decline to the current day conditions, given Titan's non-zero eccentricity. With an improved knowledge of Titan's surface topography from Cassini, such an endeavor is both possible and worthwhile for the testing of our model. Tidal modeling with the polar ocean assumption may also explore the production of volatile materials and subsequent transportation to and deposition of thick polar deposits.

Theoretical modeling of the volatile budgets necessary for the development of the empty and filled depressions may place limits on the amount of dissolvable material needed in a matrix of both soluble and insoluble components. This would include an investigation of volatile inventories, volatile losses and transformations, and the present inventory of volatiles (including, e.g., tholins). Quantitative modeling of landform evolution by simulation models (e.g., Howard et al., 1994) can also evaluate the process rates and varying environmental scenarios. As new topographic data become available, it would be particularly valuable to further characterize the landform relationships at the poles, and place further constrains on surface composition. Lastly, future missions to Titan could test the conceptual model by identifying layered sedimentary units. Investigating the variation (or uniformity) of the composition in any potential sedimentary layers could confirm or rule out our model.

## ACKNOWLEDGEMENTS

All data for this mapping project are located at: http://www.geomorph-sbirch.com/data-products/. SPDB, AGH, WED, JWB, EPT, and RLK were funded by a NASA Cassini Data Analysis Program: Grant NNX13AG03G, and SPDB by the NASA Earth and Space Science Fellowship Program: Grant 5-PLANET5F- 0011. DAW was funded for Titan geologic mapping under grant NNX14AT29G from the Outer Planets Research Program. This research was also supported by the Cassini-Huygens mission, a cooperative endeavor of NASA, ESA, and ASI managed by JPL/Caltech under a contract with NASA. We would like to thank two reviewers for their comments on the manuscript. We would especially like to thank Thomas Cornet for a very thorough and thoughtful review that significantly improved the manuscript. Finally, we would also like to acknowledge the entire Cassini RADAR team for the acquisition of the radar data, and Paul Corlies for comments and revisions of earlier versions of this manuscript.

FIGURES

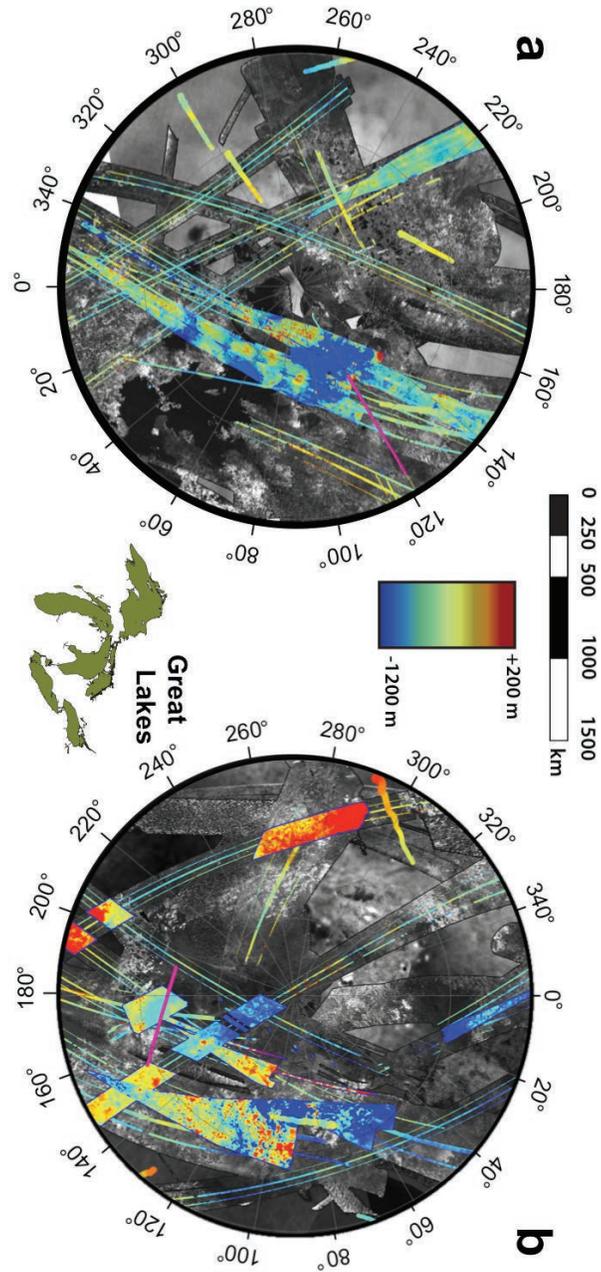

Figure 1 – All available data that was used to create our geomorphologic maps for both the (a) north and (b) south poles. Topographic data used includes DTMs, SARTopo (processed up to T92) and altimetry passes up to and including T108. Portions of altimetry passes used in Fig. 6 (T91 in the north and T49 in the south) are highlighted in pink. Great Lakes are shown for scale.

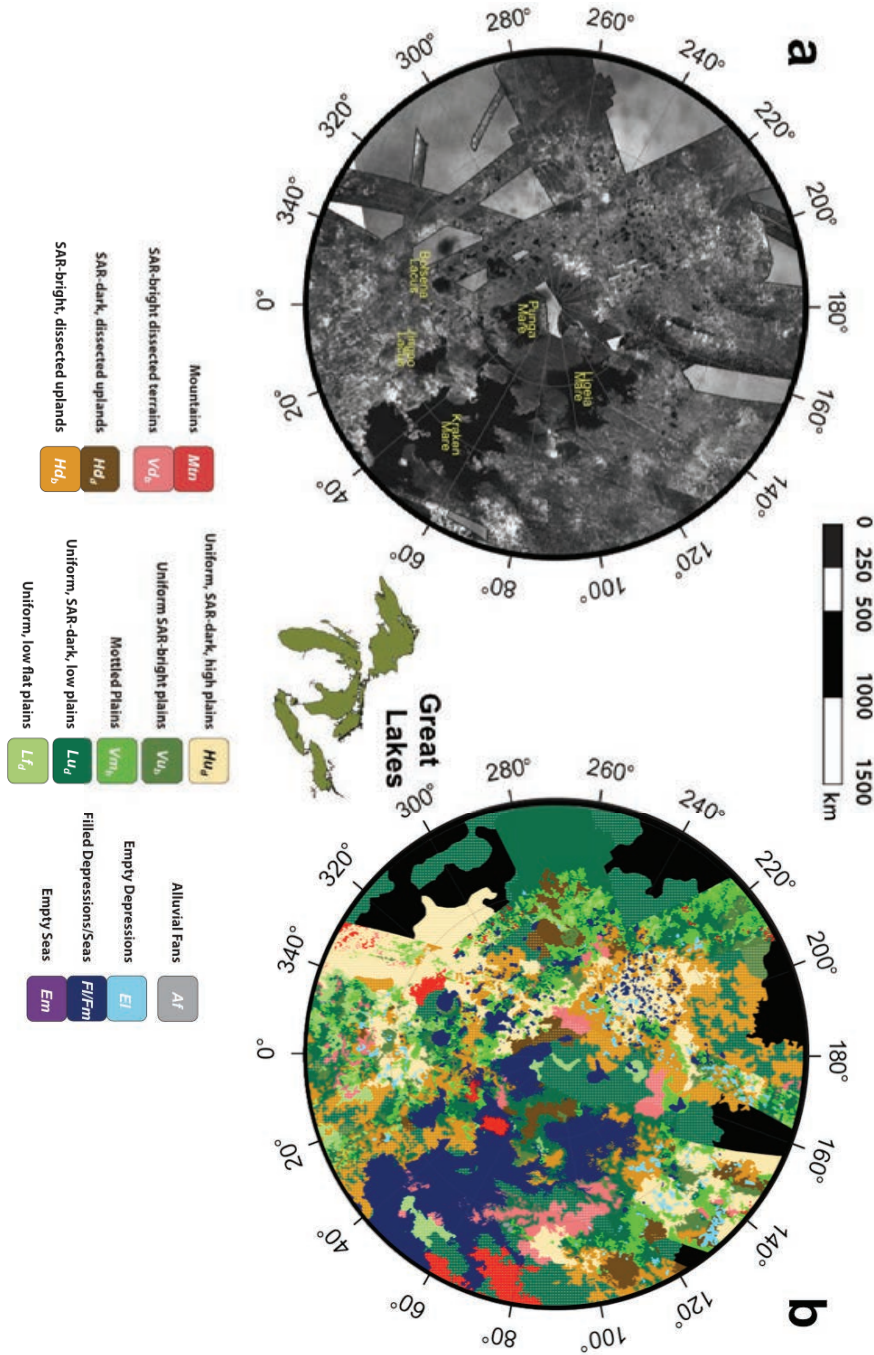

Figure 2 – (a) SAR mosaic overlain on ISS imagery of Titan's north pole. The geographic locations of Ligeia Mare, Kraken Mare, Punga Mare, Jingpo Lacus and Bolsena Lacus are shown for reference; (b) Geomorphologic map of the region produced using the SAR and ISS imagery. Areas where high resolution SAR was unavailable are shown with white or gray dots. Mapping in these regions used HiSAR and/or ISS datasets instead. Topography over these regions relied on altimetry where available. Black regions were unmapped. Great Lakes are shown for scale. A legend for the unit names and color associations is shown underneath the mapping frames.

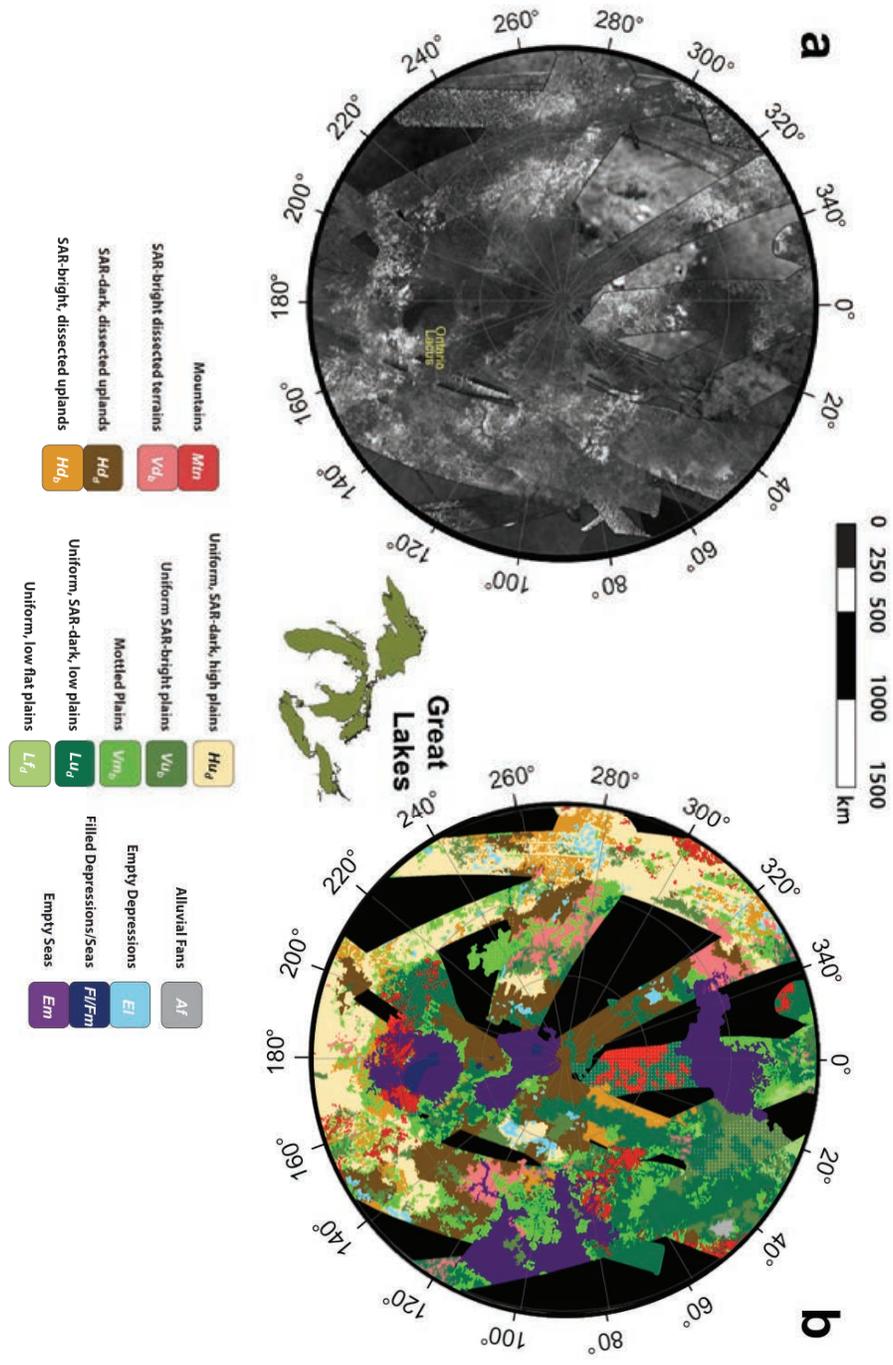

Figure 3 – (a) SAR mosaic overlain on ISS imagery of Titan's south pole. The geographic location of Ontario Lacus is shown for reference; (b) Geomorphologic map of the region produced using the SAR and ISS imagery. Areas where high resolution SAR was unavailable are shown with white or gray dots as in Fig. 2b. Black regions were unmapped. Great Lakes are shown for scale. A legend for the unit names and color associations is shown underneath the mapping frames.

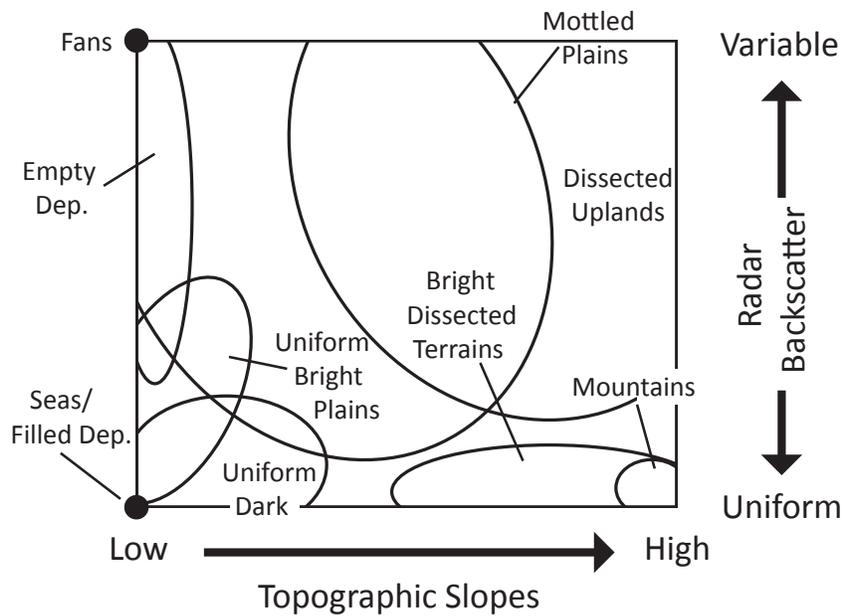

Figure 4 – Qualitative description of mapping units according to their radar texture. Properties of each unit are listed in Table 1. A variable SAR backscatter could result from small-scale roughness, dielectric constant variabilities and/or volume scattering. Units with varying appearances in multiple SAR images will appear at the top of the plot. If a unit is topographically flat but includes numerous scatterers on the surface, like the fans, then it appears in the top left. Units with high slopes are likely to be SAR-bright due to large-scale facets of the surface orientated towards the spacecraft and will be located at the bottom right of the figure.

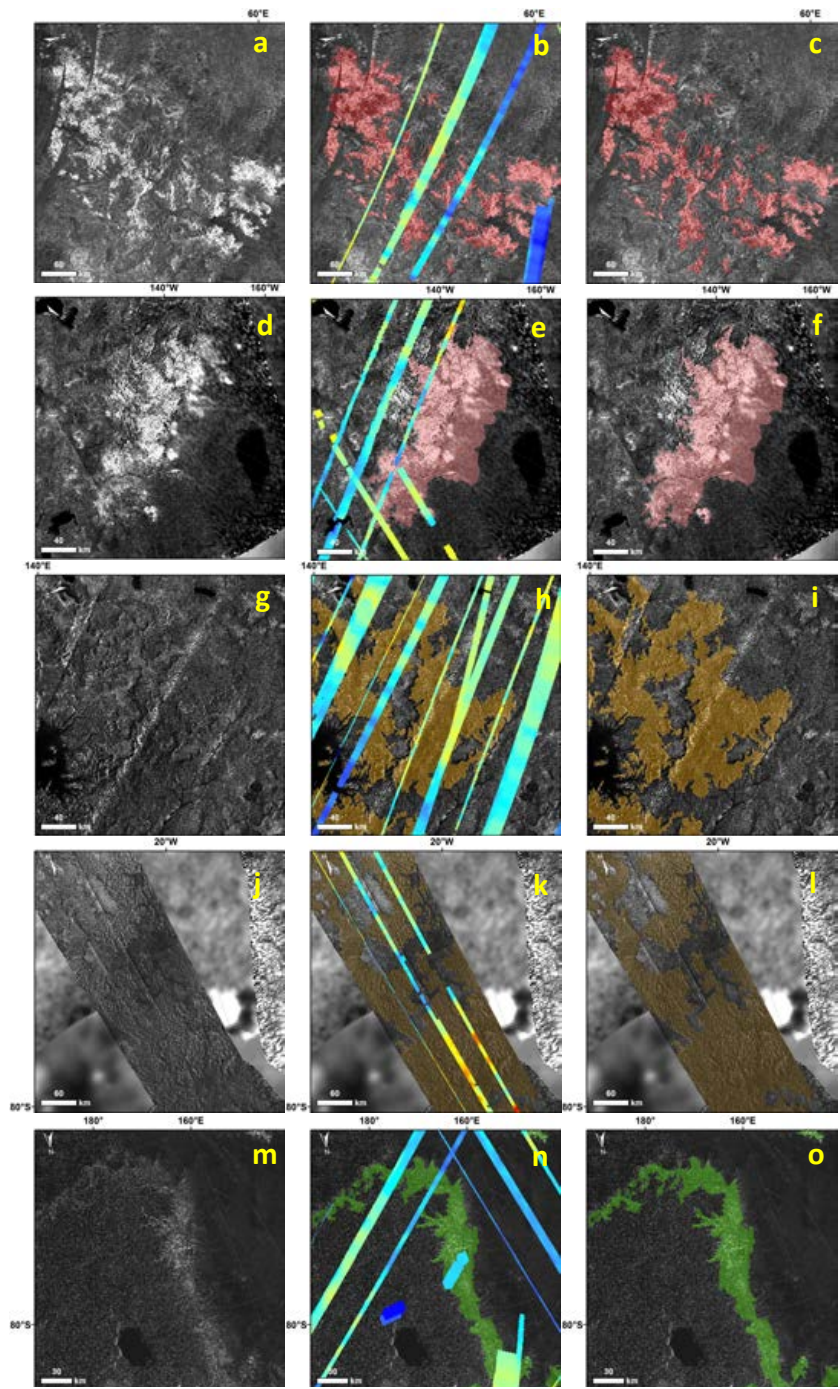

Figure 5 – Variable backscatter terrains, with the SAR image on the left, mapped image in the middle with topography overlain, and the mapped unit on the right. The scale for topography is as shown in Fig. 1; – (a/b/c) Mountains (*Mtn*) at the south; (d/e/f) SAR-bright, dissected terrains (*Vd_b*), similar appearance to mountains yet topographically depressed; (g/h/i) SAR-bright, dissected uplands (*Hd_b*) showing a varied brightness. This unit is distinguishable from the mottled plains in both their scarp-like perimeters and high relief; (j/k/l) SAR-dark, dissected uplands (*Hd_d*), part of the Sikun Labyrinthus region. Organized channel structure is evident, while the relief is high; (m/n/o) Mottled Plains (*Vm_b*) around the border of a putative empty sea.

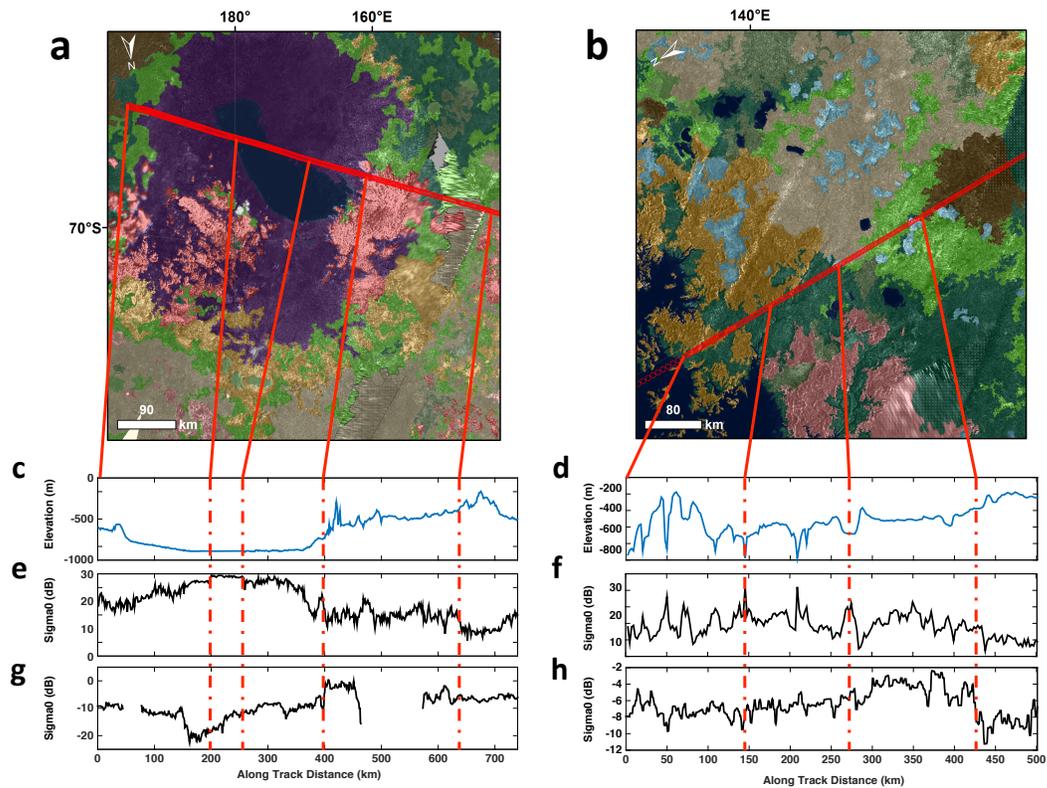

Figure 6 – (a/c/e/g) SAR image of the Ontario Lacus region, a filled depression, situated at the bottom of an empty sea. The basin itself is surrounded by mountains. Mapped unit transparencies are overlain; (c) The mountains appear with the highest relief in the image; (e) the returned altimetry signal from T48 is low compared to the specular reflection from the very flat liquid surface and SAR-dark basin floor; (g) The normalized radar backscatter ($\sigma_0$) in dB for the off-nadir SAR averaged within each footprint of the T48 altimetry signal.

(b/d/f/h) Region south of Ligeia Mare showing large valleys draining into the sea. Mapped units with transparencies are overlain. (d) Topography increases moving away from the sea into the surrounding dissected uplands; The backscatter over the SAR-dark, dissected uplands is low in both the SAR image (h) and the T91 altimetry profile (f). Topography in both cases was measured using the 1$^{st}$ moment of the altimetry signal, while the backscatter is reported in dB.

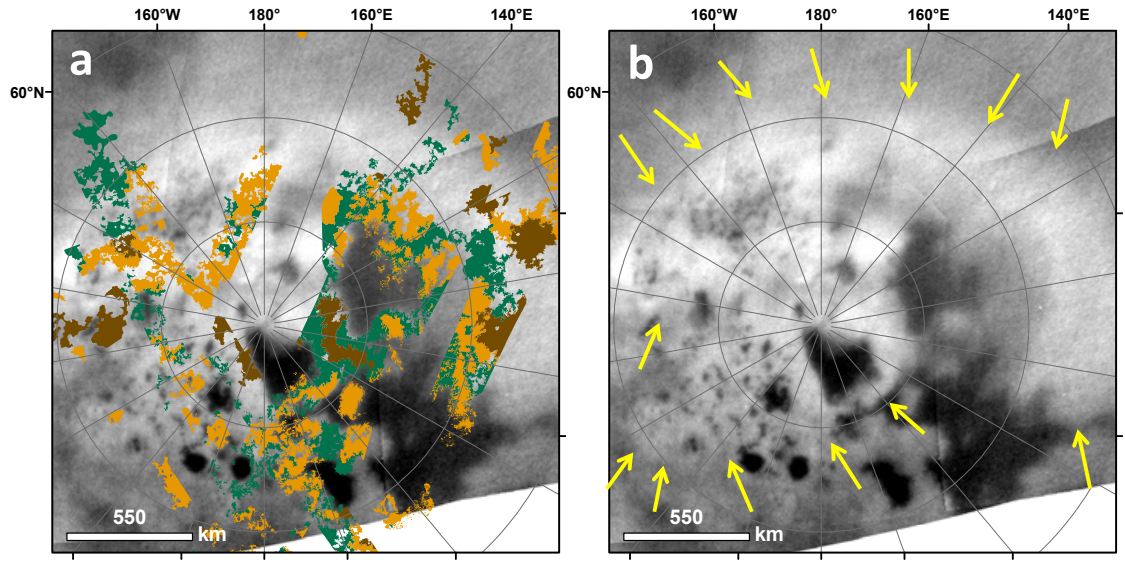

Figure 7 – (a) Mapping units created using the highest resolution SAR imagery are overlain on ISS imagery. Units that appear to correlate with ISS bright terrain are the dissected uplands ($Hd_d$ and $Hd_b$), and the SAR-dark, low plains ($Lu_d$), which would be consistent with similar compositions for the three unit types.

(b) ISS image showing what appears to be a polar deposit encompasses the north polar region and terrains extending down to ~62° N. Yellow arrows denote approximate boundaries of the unit. The high albedo ISS terrain is absent for longitudes between -20° W and 70° E, at latitudes above 60° N.

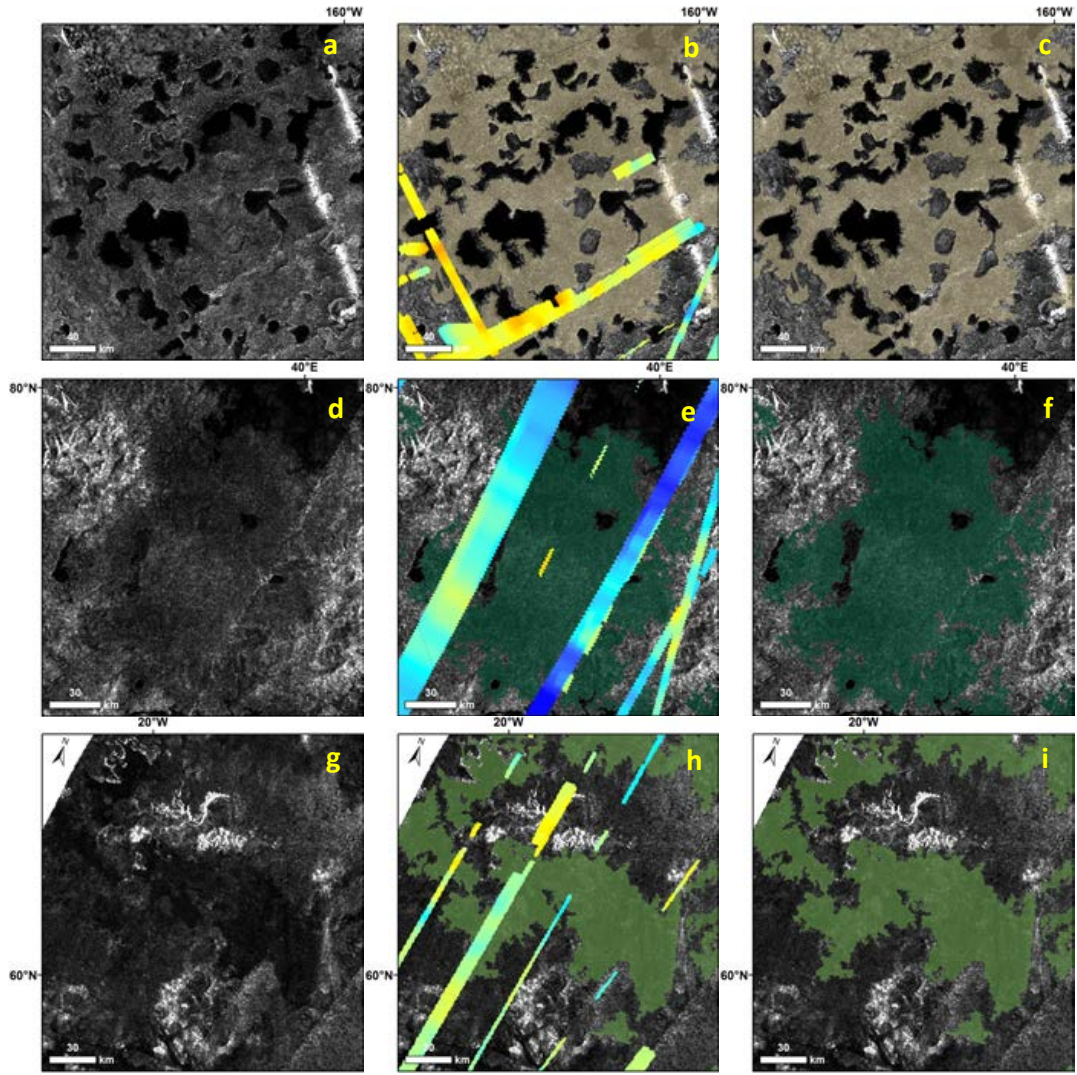

Figure 8 – Uniformly SAR-dark units, with the SAR image on the left, mapped image in the middle with topography overlain, and the mapped unit on the right. The scale for topography is as shown in Fig. 1; (a/b/c) SAR-dark, high plains ($Hu_d$). Numerous depressions are seen in the region. (d/e/f) SAR-dark, low plains ($Lu_d$) in the region between Kraken Mare and Jingpo Lacus. (g/h/i) Low flat plains ($Lf_d$) south of Bolsena Lacus. The unit is uniformly dark, with only liquids and dune materials having lower backscatters. These regions are also topographically flat over the entire spatial dimension of the unit. All colors are as in Figs. 2/3, with transparencies to highlight the units' morphologies.

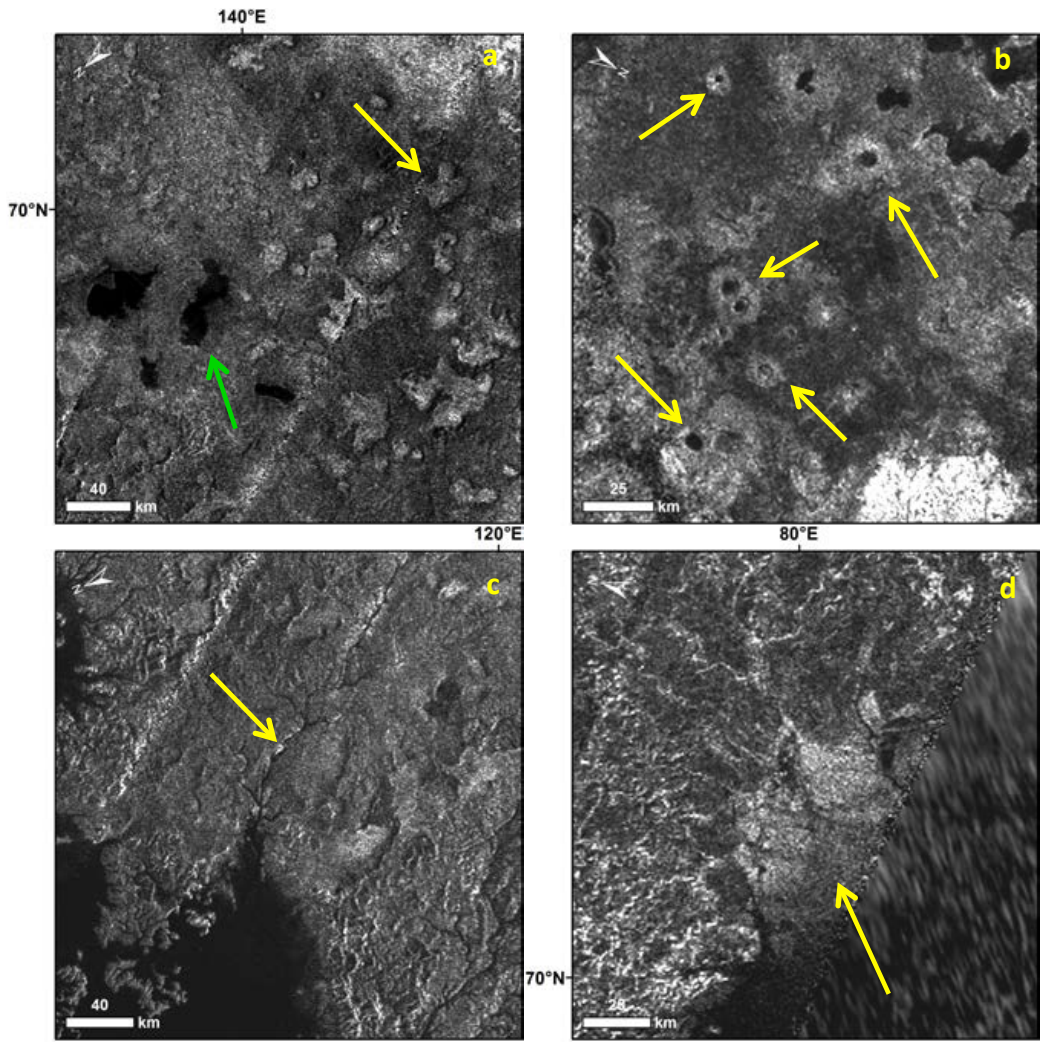

Figure 9 – Embedded units. (a) Empty (yellow arrow) and filled (green arrow) depressions (*El/Fl*) south of Ligeia Mare. The morphologies of their perimeters are similar, and both are embedded in a SAR-dark material. The floors of the empty depressions appear bright to their surroundings in this image, though such an appearance is not a requirement; (b) Circular depressions (marked by yellow arrows) appear with a much more regular perimeter surrounded by a SAR-bright mound; (c) Fluvial valley, Vid Flumina, exhibiting a dendritic pattern. Recent altimetry analysis suggests this network to be incised ~300 m into the surrounding SAR-dark plain ($Lu_d$). (d) Alluvial fans (*Af*) along the southern perimeter of Kraken Mare. These fans overlap to form a bajada, and exhibit classic alluvial fan morphologies.

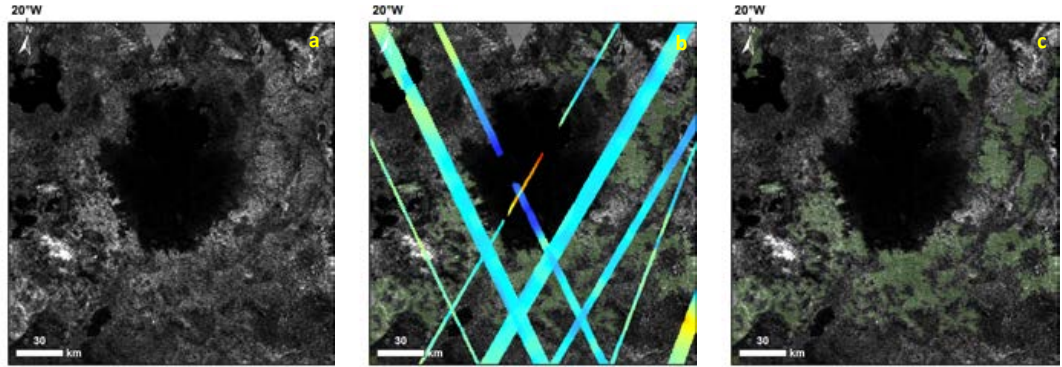

Figure 10 – Uniform SAR-bright plains south of Bolsena Lacus, with the SAR image on the left, mapped image in the middle with topography overlain, and the mapped unit on the right. The scale for topography is as shown in Fig. 1. The unit has a uniform backscatter at our mapping resolution and is topographically flat across the unit. The increased, but also uniformity, in the backscatter received from the uniform bright plains distinguishes this unit from the mottled plains.

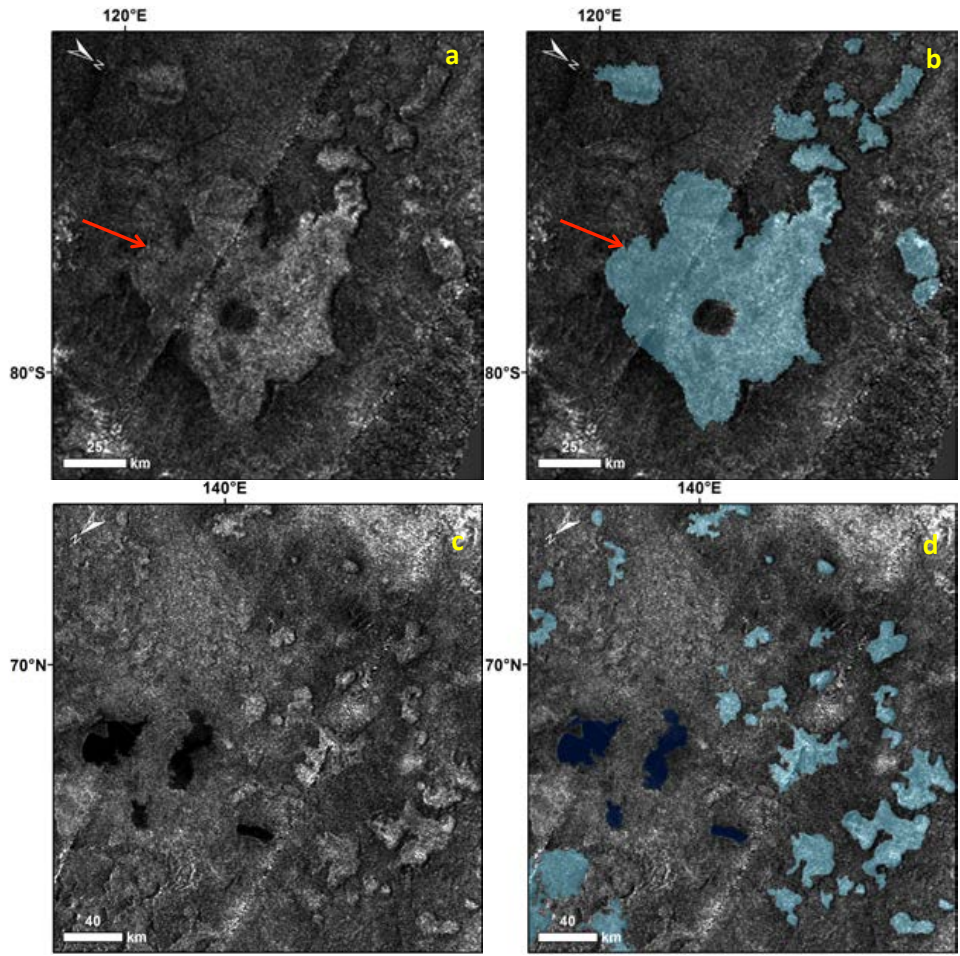

Figure 11 – Adapted from Hayes (2016); Top (a/b): Breached empty depressions (*El*) at the south show evidence for the progressive growth of multiple depressions into a single large depression. A valley network on the left is encroaching upon the depression(s), presumably leading to a breach (red arrows) and faster growth.
Bottom (c/d): Filled (*Fl*) and empty (*El*) depressions at the north. These depressions are embedded within the SAR-dark, high plains ($Hu_d$), and are bounded by the SAR-bright, dissected uplands ($Hd_b$). The empty depressions are situated in a local topographic high.

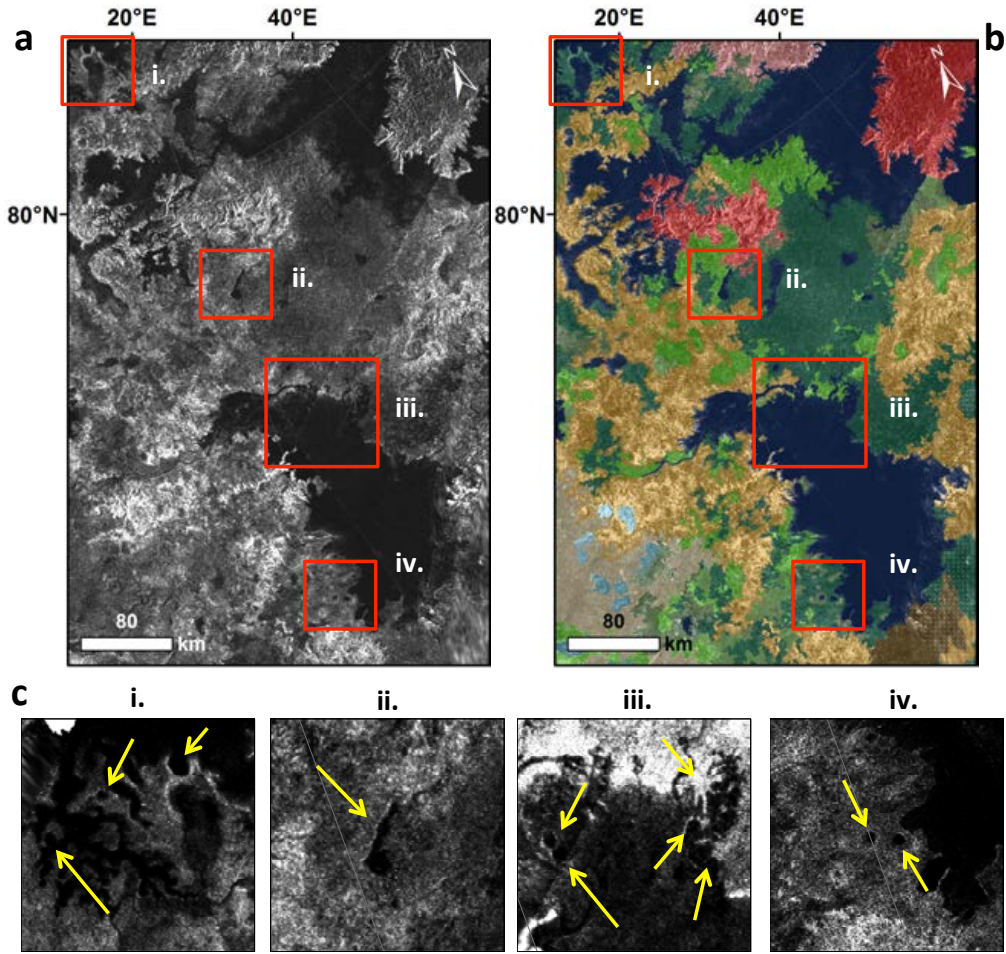

Figure 12 – (a) A region of particular interest that includes Jingpo Lacus and the connection between Kraken and Punga Mare; (b) Corresponding geomorphologic map of the region with units color coded as in Figs. 2b/3b. Red boxes highlight features in bottom panels; (c) Zoomed in panels of highlighted circular depressions are common around the shorelines of the seas (panels i and iv) and seem to elongate and coalesce into a single chain in many locations (panel ii). Circular depressions also occur at the bottom of Jingpo Lacus (iii), suggesting that the substrate on the seafloors of large liquid bodies is similar to the substrate in which smaller depressions form. Features of interest in (c) are marked with yellow arrows, with each panel 100 km across. We vary the stretch on panel iii to highlight the features of interest.

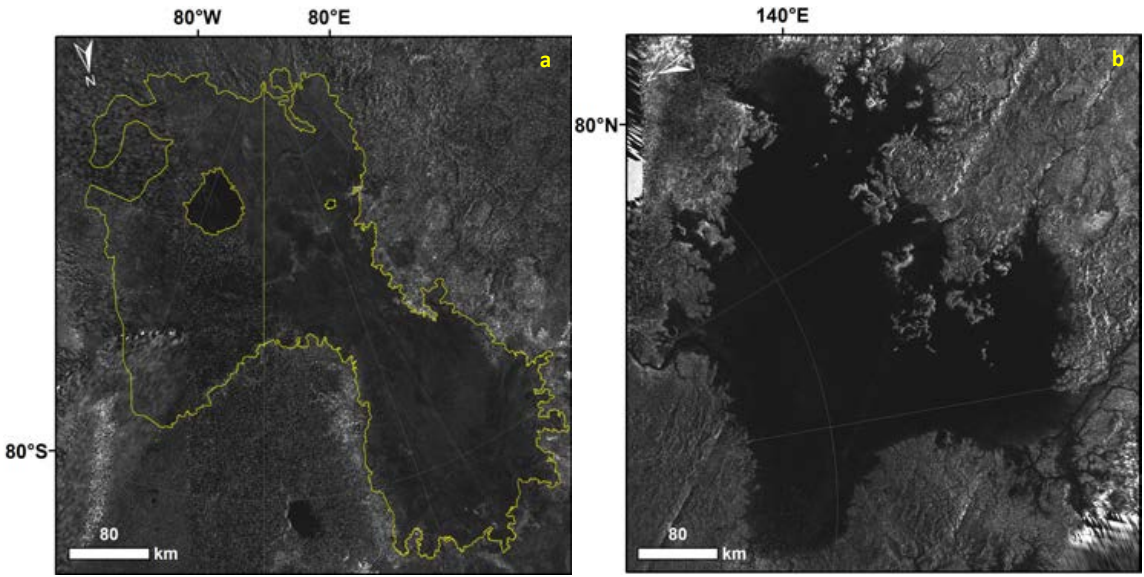

Figure 13 – (a) Empty sea (*Em*) around Titan's south pole with the boundary marked in yellow. These basins occupy the lowest elevations at the south. Evidence of valley incision is seen along their perimeters; (b) Filled sea (Ligeia Mare) at the north shows a similar shoreline morphology to the southern empty sea. The largest valley networks drain into the seas at the north, while the southern empty seas also show evidence for large drainage systems.

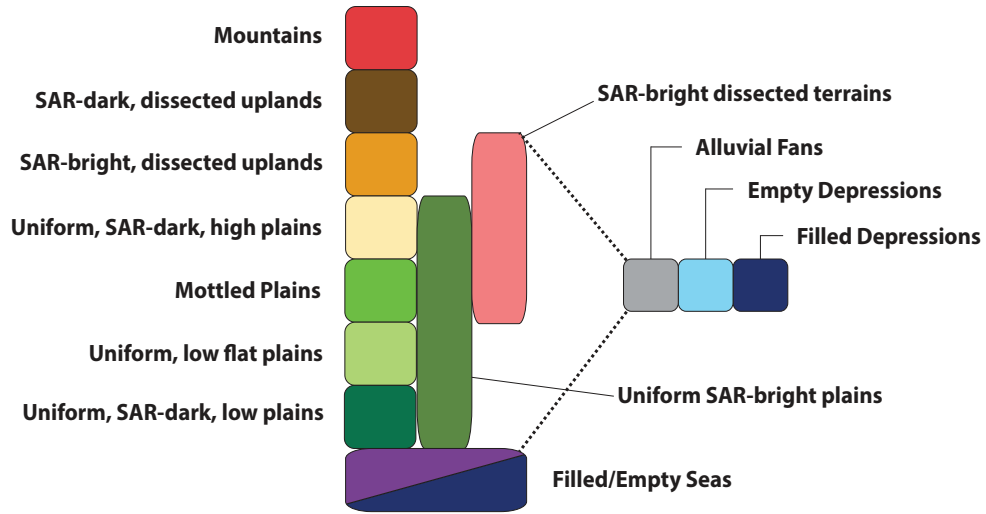

Figure 14 – Topographic relations column for mapped morphologic units. Colors are as in Figs. 2/3, and individual units are described in Section 3. The units in the column are stacked by elevation of relative occurrence. Mountains (*Mtn*) define the highest unit, and successively lower units are defined as uplands and plains. Note that this is not a stratigraphic column.

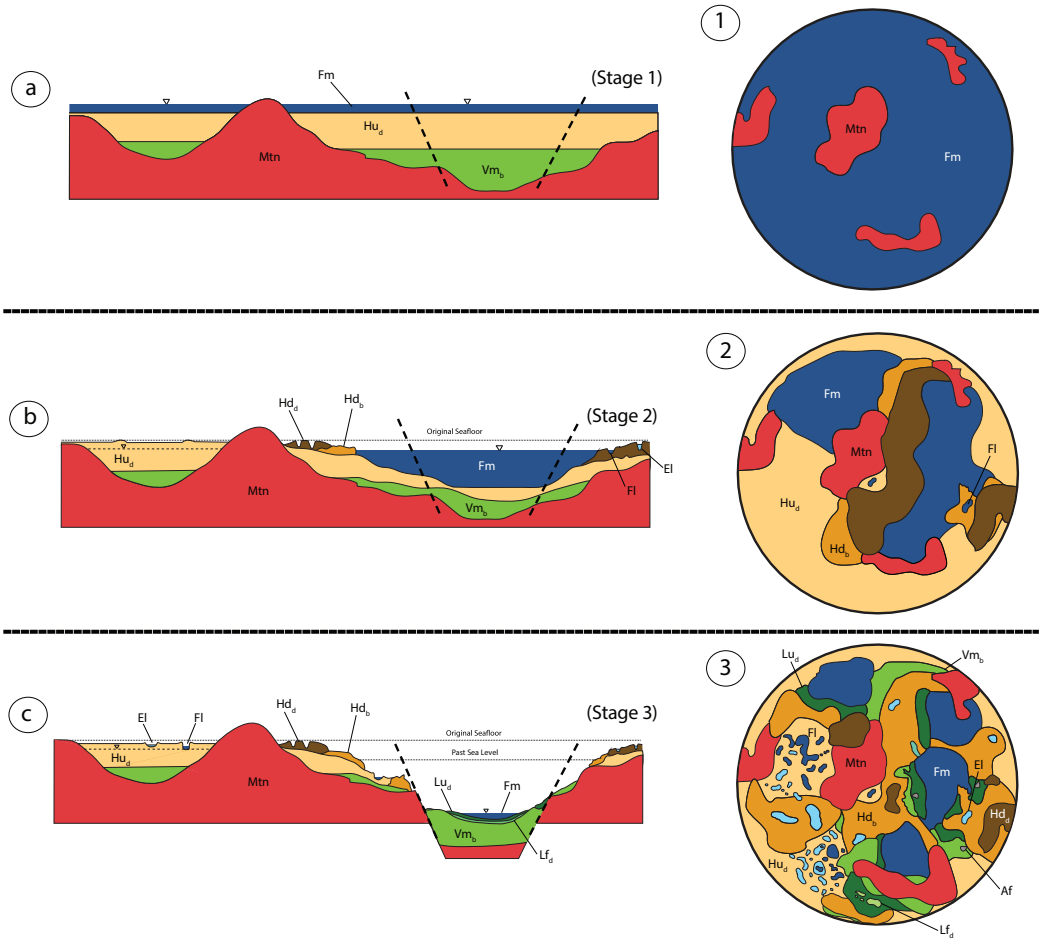

Figure 15 – Cartoon model described in Section 5. Color scheme as in Fig. 14. Relief and sediment thicknesses are exaggerated in all panels; Top: (1) Primordial, water-ice rich bedrock (*Mtn* and *Vd$_b$*) is overlain by a large methane-ethane ocean. Sediments eroded from the water-ice regolith along with precipitated atmospheric organics (*Hu$_d$*) begin to accumulate at the bottom. Associated cross-sectional view (a) is shown with only a few areas not submerged at the onset of Stage 1.

Middle: (2/b) During Stage 2, the ocean retreats to multiple large seas, with exposed sea-floor material now susceptible to erosion. The dissected uplands (*Hd$_d$* and *Hd$_b$*) are left as remnant highs. Depressions (*Fl/El*) begin to form, where initially a larger fraction of them are filled compared to the current day.

Bottom: (3) Stage 3 with a polar view of the current north polar region. Smaller seas (*Fm/Em*) are isolated from the small depressions. More empty depressions are also forming, where in previous epochs (eg. Panel 2/b) they were filled with fluids. A current day cross-sectional representation of both the north and south is shown in panel (c). Normal faulting (dashed vertical lines) is shown to illustrate vertical crustal motions and the observed stepped topography.

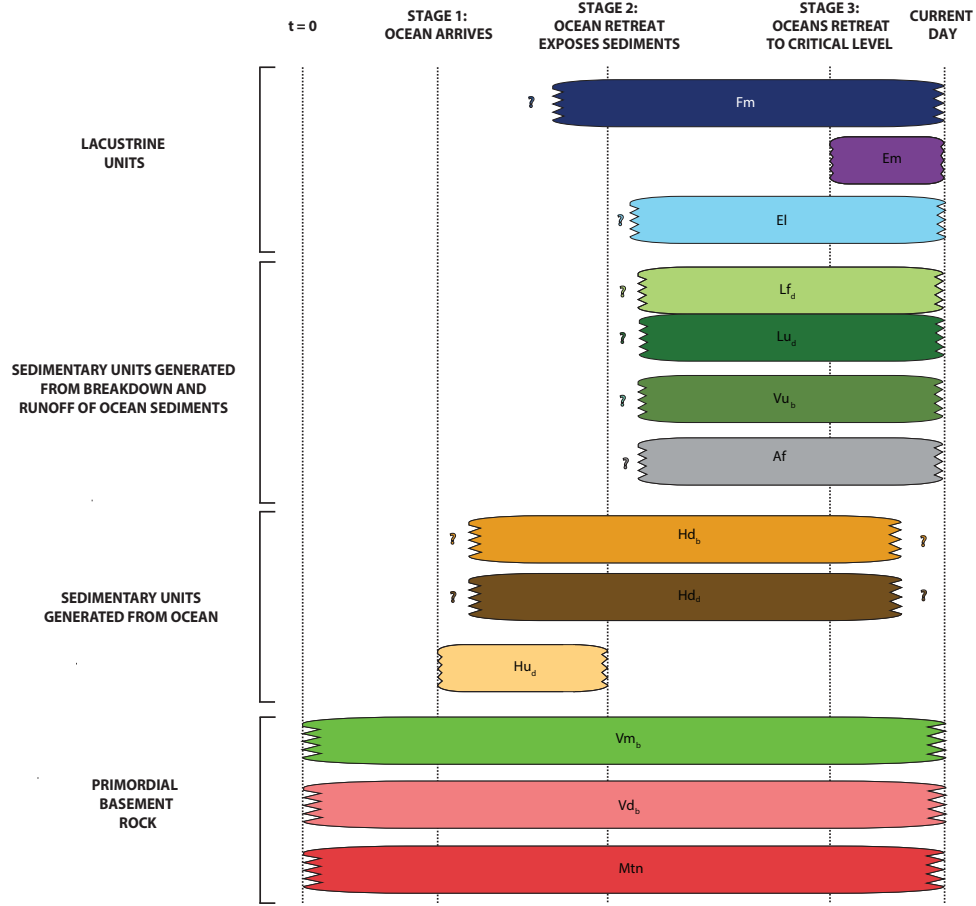

Figure 16 –. Stratigraphic column associated with Fig. 15 (Colors are as in Fig. 14) following the stages of evolution discussed in Section 5 shown at the left with dotted lines. Prior to any ocean arrival we assume Titan's surface to be dominated by a water-ice rich regolith. Large inventories of methane arise, forming a polar ocean where sediments accumulate at the bottom. As the ocean retreats, sediments are eroded and reprocessed into different units, and more empty depressions begin to appear in the landscape. Currently, as fluids are exchanged between poles, one pole becomes entirely empty while the other contains all the remaining surficial liquid volumes.

| Unit Name | Radar Scattering | Morphology & Topography | ISS Brightness | Previous Mapping Name(s) | Fractional Mapped Area (North /South) |
|---|---|---|---|---|---|
| Mountains (Mtn) | High $\sigma_0$ SAR, low $\sigma_0$ alt., bright-dark pairing SAR | Highest relief; high topographic slopes; highly dissected; no lake depressions | Medium Dark | Williams+2011: Rough Highland Material Moore+2014: Dissected Plateaus & Crenulated Terrain | (4.2% / 3.8%) |
| SAR-bright dissected terrains (Vd) | High $\sigma_0$ SAR, low $\sigma_0$ alt., bright-dark pairing SAR | Variable relief; high topographic slopes; highly dissected; no lake depressions | Medium Dark | Lopes+2010/2016: Hummocky/Mountainous Terrain | (3.3% / 3.4%) |
| SAR-dark dissected uplands (Hd) | Low $\sigma_0$ SAR, low $\sigma_0$ alt. | 2nd highest relief; high dissection; few lake depressions, scarp-like perimeters | Bright | Malaska-2016: Labyrinth Terrain Lopes+2016: Labyrinth Terrain | (3.2% / 9.5%) |
| SAR-bright dissected uplands (Hd) | High $\sigma_0$ SAR, appearance varies between SAR swaths | High relief; high dissection; numerous lake depressions, scarp-like perimeters | Bright | Malaska-2016: Labyrinth Terrain & Scalloped Terrain Lopes+2010: Hummocky/Mountainous | (13.7% / 4.3%) |
| Mottled Plains (Vm) | Variable $\sigma_0$ SAR | Variable relief; low, undulating local slopes; highly dissected appearance | No Clear Signal | Lopes+2010: Mottled Plains Moore+2014: Dissected Mottled Terrain | (6.7% / 9.8%) |
| Uniform, SAR-dark, high plains (Hu) | Uniformly low $\sigma_0$ SAR | Topographically high, low, undulating local slopes; uniform, undissected appearance; contain few fluvial valleys and most lake depressions | Medium Dark | N/A | (11.6% / 13%) |
| Uniform, SAR-dark, low plains (Lu) | Uniformly low $\sigma_0$ SAR | Topographically low, low, undulating local slopes; uniform, undissected appearance; contain fluvial valleys and a few lake depressions | Bright | N/A | (26% / 14.3%) |
| Uniform, low, flat plains (Lf) | Lowest $\sigma_0$ SAR besides liquids bodies | Topographically low, low, flat, local slopes; uniform, undissected appearance; very SAR-dark, liquid-like appearance | Very Dark | N/A | (3.3% / 0.9%) |
| Uniform, SAR-bright plains (Wu) | Uniformly high $\sigma_0$ SAR | Variable elevations; low, flat, local slopes; uniform, undissected appearance | No Clear Signal | Malaska-2016: Variable Featured Plains | (2.9% / 5.2%) |
| Empty/Filled Depressions (El/Fl) | El: High $\sigma_0$ SAR and alt. Fl: Lowest $\sigma_0$ SAR, high $\sigma_0$ alt. | Topographically enclosed depressions; steep sided; flat floored; few channels entering; can appear liquid filled or empty | El: No Clear Signal Fl: Very Dark | Hayes+2008: Empty/Filled/ Granular Lakes | El - (1.8% / 1.5%) Fl - (2.7% / 0.4%) |
| Empty/Filled Seas (Em/Fm) | Em: Low $\sigma_0$ SAR Fm: Low $\sigma_0$ SAR, highest $\sigma_0$ alt. | Broad depressions; termination point for largest channels; topographically enclosed; lowest regions at each pole | Em: Dark Fm: Very Dark | Hayes+2008: Mare (Kraken Mare, Punga Mare & Ligeia Mare) | Em- (0.0% / 10.4%) Fm - (11.2% / 0.0%) |
| Fluvial Valleys | Can have high or low $\sigma_0$ SAR, filled valleys have high $\sigma_0$ alt. | Linear features; Can appear SAR-dark or bright; May terminate in liquid body or alluvial fans | No Clear Signal | Burr+2013: Fluvial Features | N/A |
| Alluvial Fans (Af) | Uniformly high $\sigma_0$ SAR | Emanate from elevated topography; uniform, undissected fan-shaped morphology; lobate, undissected surfaces | No Clear Signal | Birch+2016: Alluvial Fans | (0.1% / 0.2%) |

Table 1 – Unit characteristics as described in Section 3 and Section 4. The second column describes their SAR backscatter appearance. A unit's altimetry (alt.) scattering behavior is also noted where it was possible to measure. $\sigma_0$ denotes the normalized radar backscatter cross-section. The third column notes the unit's morphologic appearance, topographic expression and presence of embedded units where applicable. The remaining three columns denote the unit's: ISS brightness, mapping names used by authors in previous works, and the percentage of area mapped at both poles.